\documentclass[fleqn,usenatbib]{mnras}

\usepackage[T1]{fontenc}

\DeclareRobustCommand{\VAN}[3]{#2}
\let\VANthebibliography\thebibliography
\def\thebibliography{\DeclareRobustCommand{\VAN}[3]{##3}\VANthebibliography}

\usepackage{graphicx}	
\usepackage{amsmath}	
\usepackage{subfig}

\title[Ceres - transport to NEO region via 8:3 MMR]{Possibility of transporting material from Ceres to NEO region via 8:3 MMR with Jupiter}

\author[M. Kov\'{a}\v{c}ov\'{a} et al.]{
M. Kov\'{a}\v{c}ov\'{a},$^{1}$\thanks{E-mail: kovacova308@uniba.sk}
L. Korno\v{s},$^{1}$
and P. Matlovi\v{c}$^{1}$
\\
$^{1}$Faculty of Mathematics, Physics and Informatics; Comenius University in Bratislava; Slovak Republic\\
}

\date{Accepted XXX. Received YYY; in original form ZZZ}

\pubyear{2021}

\begin{document}
\label{firstpage}
\pagerange{\pageref{firstpage}--\pageref{lastpage}}
\maketitle

\begin{abstract}
In this work we investigate the possibility of transporting material to the NEO region via the 8:3 MMR with Jupiter, potentially even material released from the dwarf planet Ceres. By applying the FLI map method to the 8:3 MMR region in the orbital plane of Ceres, we were able to distinguish between stable and unstable orbits. Subsequently, based on the FLI maps (for mean anomaly $M=60^\circ$ and also $M=30^\circ$), 500 of the most stable and 500 of the most unstable particles were integrated for $15\,Myr$ for each map. Long-term integration in the case of $M=60^\circ$ showed that most of the stable particles evolved, in general, in uneventful ways with only 0.8\% of particles reaching the limit of q $\leq$ 1.3 $AU$. However, in the case of $M=30^\circ$, a stable evolution was not confirmed. Over 40\% of particles reached orbits with q $\leq$  1.3 $AU$ and numerous particles were ejected to hyperbolic orbits or orbits with a > 100 $AU$. The results for stable particles indicate that short-term FLI maps are more suitable for finding chaotic orbits, than for detecting the stable ones. A rough estimate shows that it is possible for material released from Ceres to get to the region of 8:3 MMR with Jupiter. A long-term integration of unstable particles in both cases showed that transportation of material via 8:3 MMR close to the Earth is possible.
\end{abstract}




\section{Introduction}
Interplanetary material in the form of interplanetary dust, meteoroids (bodies with size $30\,\mu m - 1\,m$) and even larger objects is transported close to the orbit of the Earth from different parts of the Solar System. It is estimated that several tens of tons of material enter the atmosphere of the Earth every day. Meteoroids represent a threat to satellites but larger objects can even threaten life on the Earth. Therefore, it is important to study where these bodies come from and how they can arrive at the Earth. The original orbits of such objects are affected by gravitational perturbations and non-gravitational effects. 
 
Numerical integrations of particles injected into the resonances in the main asteroid belt help us to understand the importance of resonances. For example, simulations of \citet{gladm} revealed that the typical end state of particles initially placed to $\nu _6$ resonance is a collision with the Sun ($80\,\%$) or ejection on a hyperbolic orbit ($12\,\%$). With a better understanding of resonant phenomena, it is now obvious that the main asteroid belt is a significant source of near-Earth objects (NEOs). Under the action of resonances, asteroids (or meteoroids) can get on a planet-crossing orbit by increasing their eccentricities \citep{Froeschl1999DynamicalTM, Morbidelli2002}. The main sources of material transported to the NEO region are secular resonance $\nu_6$ at the inner edge of the asteroid belt and 3:1 mean motion resonance (MMR) with Jupiter \citep{bottke}. It has also been shown that the powerful resonances 5:2 and 2:1 MMR with Jupiter are not as important as sources of NEOs \citep{granvik}. The most notable resonances, such as 3:1, 5:2, 7:3, 2:1 MMRs with Jupiter, can also be spotted as gaps in the distribution of asteroids in the main asteroid belt - so-called Kirkwood gaps. In addition to these wide MMR resonances, the main asteroid belt is crossed by large numbers of thin resonances (higher-order MMRs, three-body resonances, etc.) \citep{Morbidelli2002}. In general, resonances represent a powerful mechanism for transporting material to various parts of the Solar System and are thus important in dynamical evolution.

Ceres is the most massive body in the main asteroid belt between Mars and Jupiter. This dwarf planet was investigated by NASA's Dawn mission, which was the first space mission to orbit and explore two extraterrestrial bodies - Vesta and Ceres. After exploring the asteroid Vesta, the Dawn spacecraft entered an orbit around Ceres in March 2015. Close-up views of Ceres revealed that the otherwise homogeneous surface of Ceres is covered with craters and dotted with anomalously bright spots \citep{spots}. The surface composition of Ceres shows fairly uniform and widespread distributions of NH$_4$- and Mg-phyllosilicates and carbonates \citep{surface}. The bright areas are deposits made mostly of Na- or  Mg-carbonate \citep{desanctis-spots, spots2}.

The delivery of meteoroids from Ceres might be of interest to meteor spectroscopists. Several works have now reported the detection of Na-rich meteoroids from asteroidal orbits
\citep{borovicka2005, vojacek2015, matlovic2019}. Though the increased Na content in some of these may be caused by an incomplete correction for the low excitation of Na I \citep{matlovic}, it is possible that some of these bodies reflect a real Na-rich asteroidal composition. The origin of Na-rich meteoroids in salt deposits similar to the bright spots on Ceres remains one of the speculated explanations of these unusual spectra \citep{borovicka2019}.

In this work we study the 8:3 MMR with Jupiter because of its relative proximity to the dwarf planet Ceres. The 8:3 MMR on its own is usually not treated as a potential source of NEOs. This resonance in combination with 5:2 MMR with Jupiter has also been investigated by \cite{todor} and \cite{deleon}. These authors concluded that it is possible to dynamically connect Pallas (or the Pallas family) and the asteroid Phaethon via these resonances. 
We decided to study whether the Ceres could potentially be the source of meteoroids transported close to the Earth. Hence we wanted to know if the 8:3 MMR is powerful enough to send material to the Earth, possibly material released from Ceres.

\section{Method - FLI}
The Fast Lyapunov Indicator (FLI) is a sensitive and easy to implement tool for detecting chaos in dynamical systems. It is based on monitoring of the evolution of the tangent vectors. The idea was introduced by \citet{froe} and consequently further developed, so it is possible to find various definitions of FLI in different sources. 

Let us consider the dynamical system defined by the set of differential equations:
	\begin{equation} \label{system}
	\frac{\mathrm{d} x}{\mathrm{d} t} = F (x), \quad x = (x_1, x_2, ...\, x_n) \, .
	\end{equation}
For any solution $x(t)$ with initial conditions $x(0)$ the evolution of deviation vector $v(t)$ is characterised by variational equations (i.e., linearised equations of motion that describe how the initial deviation vector $v(0)$ evolves with time):
	\begin{equation}
	\frac{\mathrm{d}v}{\mathrm{d}t} = \frac{\mathrm{d}F}{\mathrm{d}x}\,v \,.
	\end{equation}
Then one way of defining the FLI for the system (\ref{system}) for initial conditions $x(0)$ and $v(0)$ at time $t$ is \citep{lega}:
	\begin{equation}
	FLI(x(0), v(0), t) = \sup_{0\leq k \leq t} \log ||v(k)||\,,
	\end{equation}
where $||v(k)||$ represents the norm of deviation vector $v(k)$.

The FLI enables us to distinguish between chaotic and regular orbits. For regular orbits, the norm of deviation vector grows linearly in time; for chaotic ones, it grows approximately exponentially. As a result, the FLI increases almost linearly with time for chaotic orbits and approximately logarithmically for regular orbits. Therefore, the computation of FLI on a grid of initial conditions $x(0)$ for the same fixed initial deviation vector allows us to map the chaos in a dynamical system (FLI map) \citep{lega, todor2017}.

It should be remarked that FLI depends on the initial deviation vector $v(0)$ and also on the integration time $t$. In order to minimize the dependence on the choice of $v(0)$, it is suggested that the average (or maximum) of the FLI values obtained for an orthonormal basis of deviation vectors should be determined. Similarly, in the case of time, a suitable integration time should be chosen by computing the FLI for different values of time \cite{lega}.

The FLI was successfully applied in studies of the outer part of the Solar System \citep{guzzo}, the neighbourhood of the Earth \citep{rosengren}, the main belt region \citep{todornovak, todor} and even exoplanet systems \citep{schwarz}. It was also applied in completely different studies, e.g., to study intramolecular dynamics \citep{shchekinova}.

For our purposes, we also decided to use the freely available software package REBOUND to simulate the motion in the Solar System. There are several integrators available to choose from. In our work we used WHFast \citep{whfast}, MERCURIUS \citep{mercurius} and IAS15 \citep{ias15}. The REBOUND is also convenient for us because variational equations  (needed to determine the FLI value) are already implemented in the form of so-called variational particles, whose evolution is characterised by variational equations \citep{vari}.

\section{Mapping the 8:3 MMR with Jupiter}
The first step in order to study the 8:3 MMR with Jupiter ($a\sim 2.705\,AU$) was to map and visualize the chaos using the FLI indicator in the region of this resonance. What the FLI map looks like depends on the epoch for which the map is computed and for which the initial distribution of perturbing objects is defined. It should also be noted that we are trying to map 6-dimensional space (defined by all orbital elements), so we always map only a part of the space. We were interested in the dwarf planet Ceres, so the angular orbital elements (inclination, argument of perihelion and longitude of ascending node) were chosen according to Ceres values, i.e., $i=10.594^\circ$, $\omega=73.37^\circ$, $\Omega=80.306^\circ$. Therefore, if it is not specified otherwise, the FLI maps in this work are maps in the orbital plane of dwarf planet Ceres.
The idea of using the FLI map of the MMR in the orbital plane of a large asteroid for selecting particles for further analysis was used for the first time in \citet{todor}.

\begin{figure*}
\centering
  \includegraphics[width=1\textwidth]{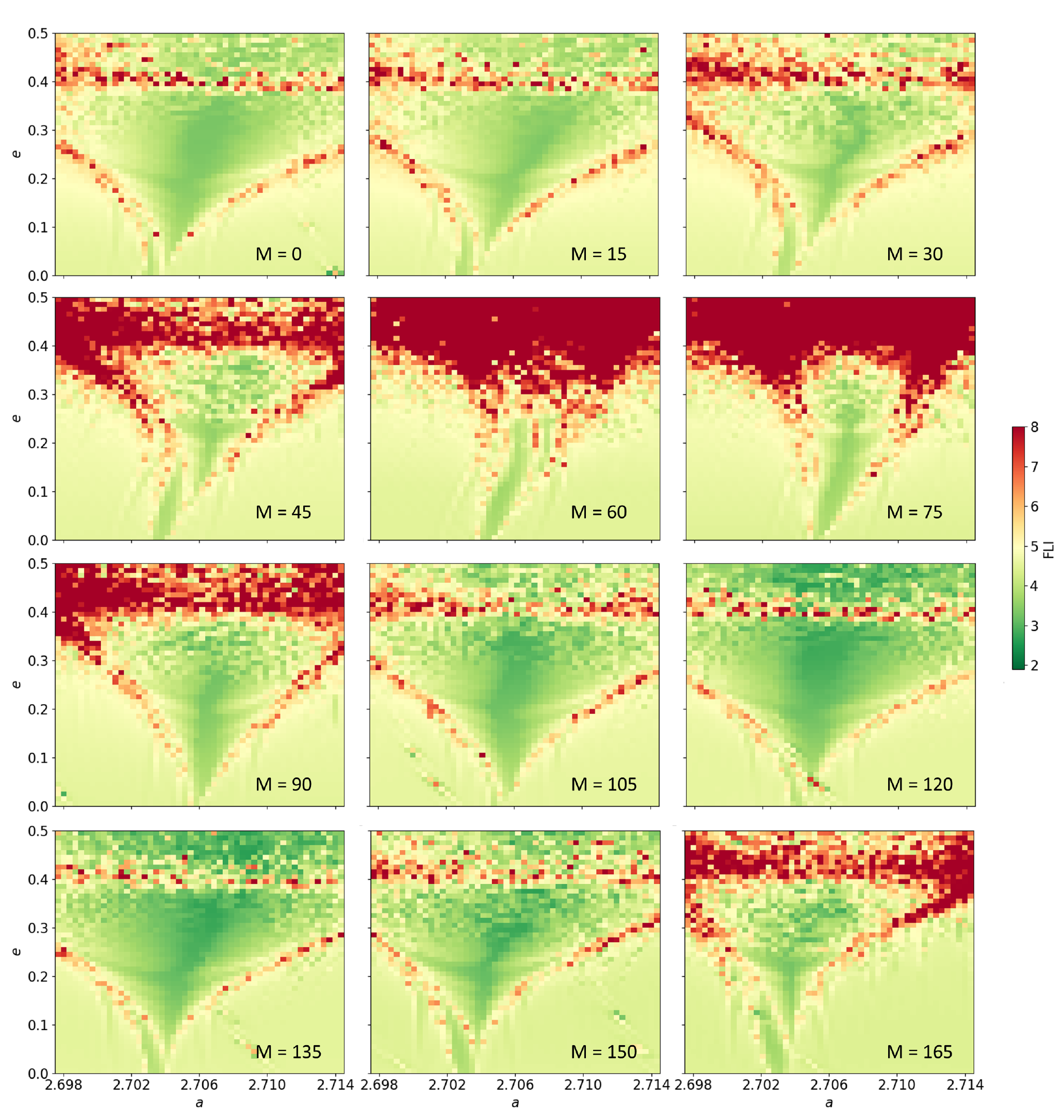}
\caption{FLI maps of the 8:3 MMR with Jupiter in orbital plane of Ceres ($i=10.594^\circ$, $\omega=73.37^\circ$, $\Omega=80.306^\circ$) for different values of mean anomaly $M \in [0^\circ, 165^\circ]$ . The maps were computed for $5\,kyr$ on a grid of $50 \times 50$ initial conditions (test particles), whereas semi-major axis $a \in \left\langle 2.6975, 2.7145 \right\rangle \,AU$ and eccentricity $e \in \left\langle 0, 0.5 \right\rangle$. Colour scale represents FLI values determined as a maximum from values obtained for six different initial deviation vectors. Stable particles with smaller FLI values are green, while the most chaotic ones, with large FLI values, are red. The width of the resonance and the amount of chaos on the maps repeat approximately after every $120^\circ$ in mean anomaly $M$.}
\label{50x50a}
\end{figure*}

\begin{figure*}
\centering
  \includegraphics[width=1\textwidth]{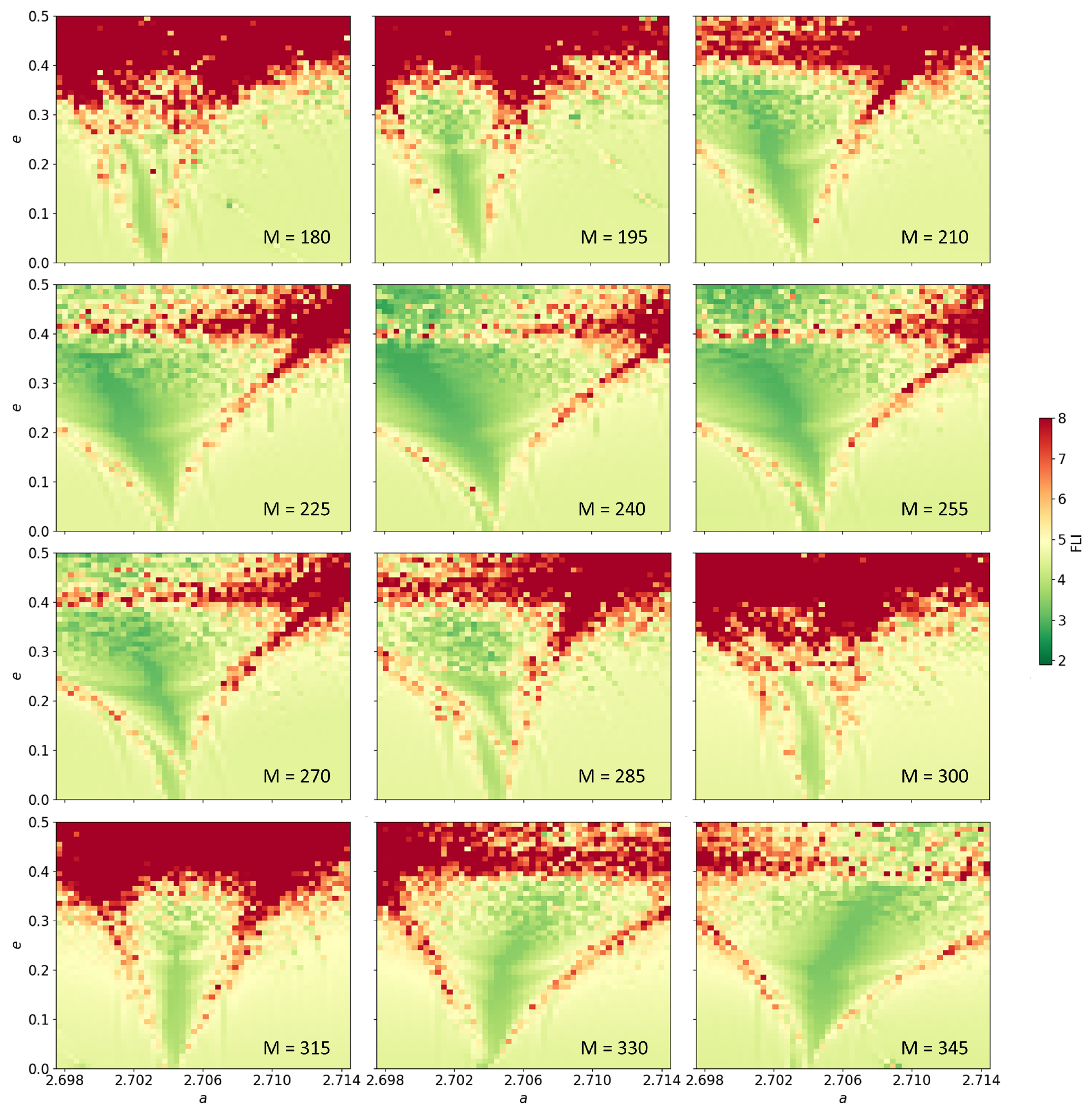}
\caption{FLI maps of the 8:3 MMR with Jupiter in orbital plane of Ceres (as Fig. \ref{50x50a}) for values of mean anomaly $M \in [180^\circ, 345^\circ]$. The width of the resonance and the amount of chaos on the maps repeat approximately after every $120^\circ$ in mean anomaly $M$.}
\label{50x50b}
\end{figure*}

Fig. \ref{50x50a}-\ref{50x50b} represent FLI maps of the 8:3 MMR region in the orbital plane of Ceres. These maps were computed for the epoch $MJD=58484.0$ (January 1 2019) for an evenly distributed mean anomaly - every $15^\circ$ in the interval $M \in [0^\circ, 360^\circ]$. Our simulation took into account the Sun and all of the planets and was integrated for $5000\,yr$ by WHFast integrator. Each map represents the grid of $50\times 50$ initial conditions. The colour scale matches the FLI value that was determined as a maximum value from the values obtained for six different initial deviation vectors. As the initial deviation vectors, we decided to use the unit vectors corresponding to varying all of the components of the position and the velocity.

Looking at the maps, it can be noticed that the map at $M=0^\circ$ is almost identical to the map at $M=120^\circ$ and very similar to the map at $M=240^\circ$. Similarly, the investigated MMR is narrowest for $M=60^\circ$, $M=180^\circ$ and $M=300^\circ$. The similarity between the maps with phase difference of $120^\circ$ basically holds for all values of $M$. This can be explained by generating the 8:3 MMR during three revolutions of Jupiter along the Sun. Moreover, we can notice a certain symmetry corresponding to the centre of the resonance ($a \sim 2.705\,AU$). For example, the resonance at $M=60^\circ$ is slightly tilted to the right, whereas the resonance at $M=180^\circ$ and $M=300^\circ$ is slightly tilted to the left.
These maps also show us that the degree of chaos is very local thing. It does not only depend on the shape and the orientation of the orbit in space, but also on the initial position of the investigated body on the orbit. And, therefore, the resolution of $50\times 50$ initial conditions is not enough for further analysis.

\section{Higher resolution of maps}

\begin{figure*}
\begin{center}
	\subfloat[]{\includegraphics[width=0.4\textwidth]{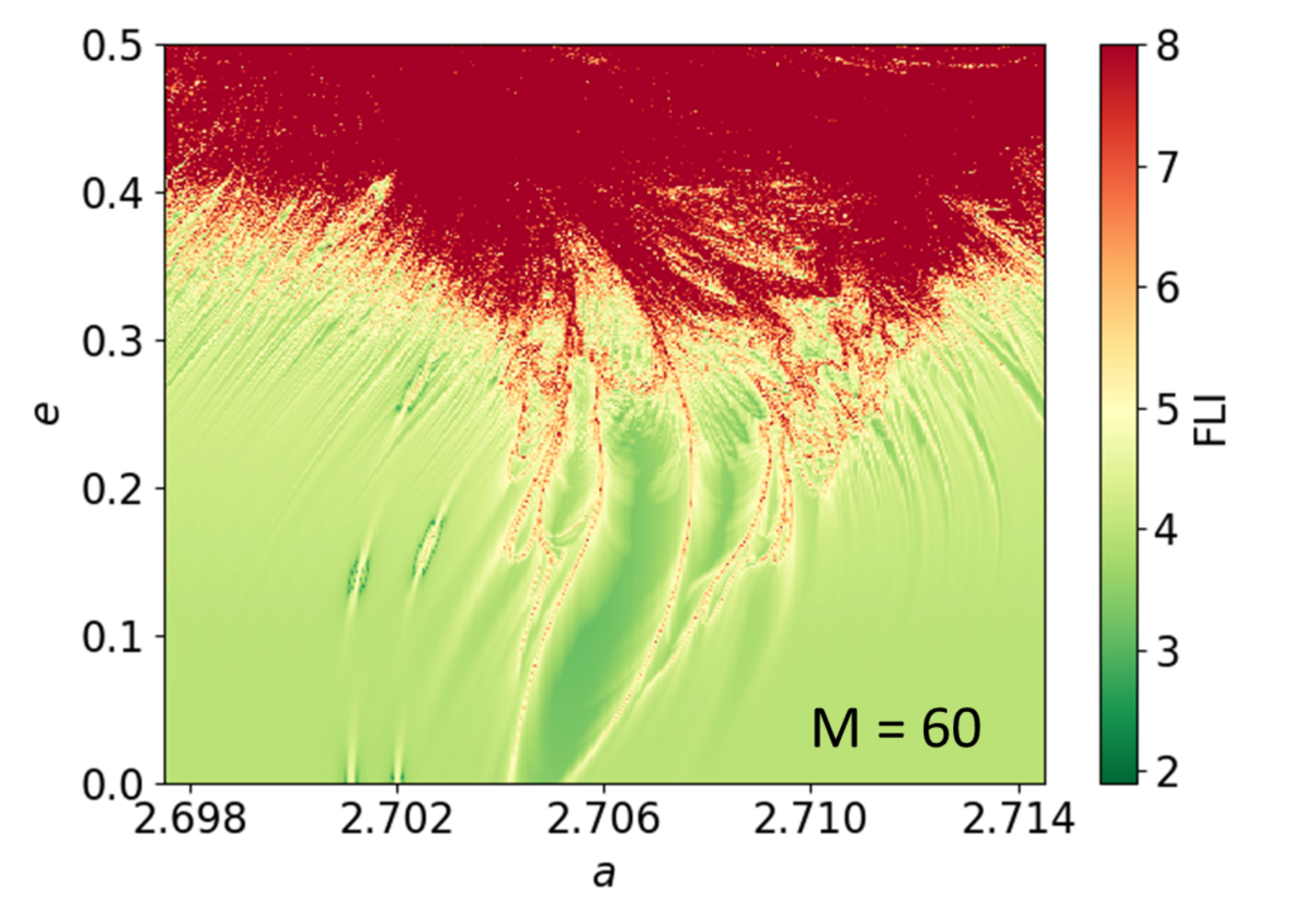}}
	\subfloat[]{\includegraphics[width=0.4\textwidth]{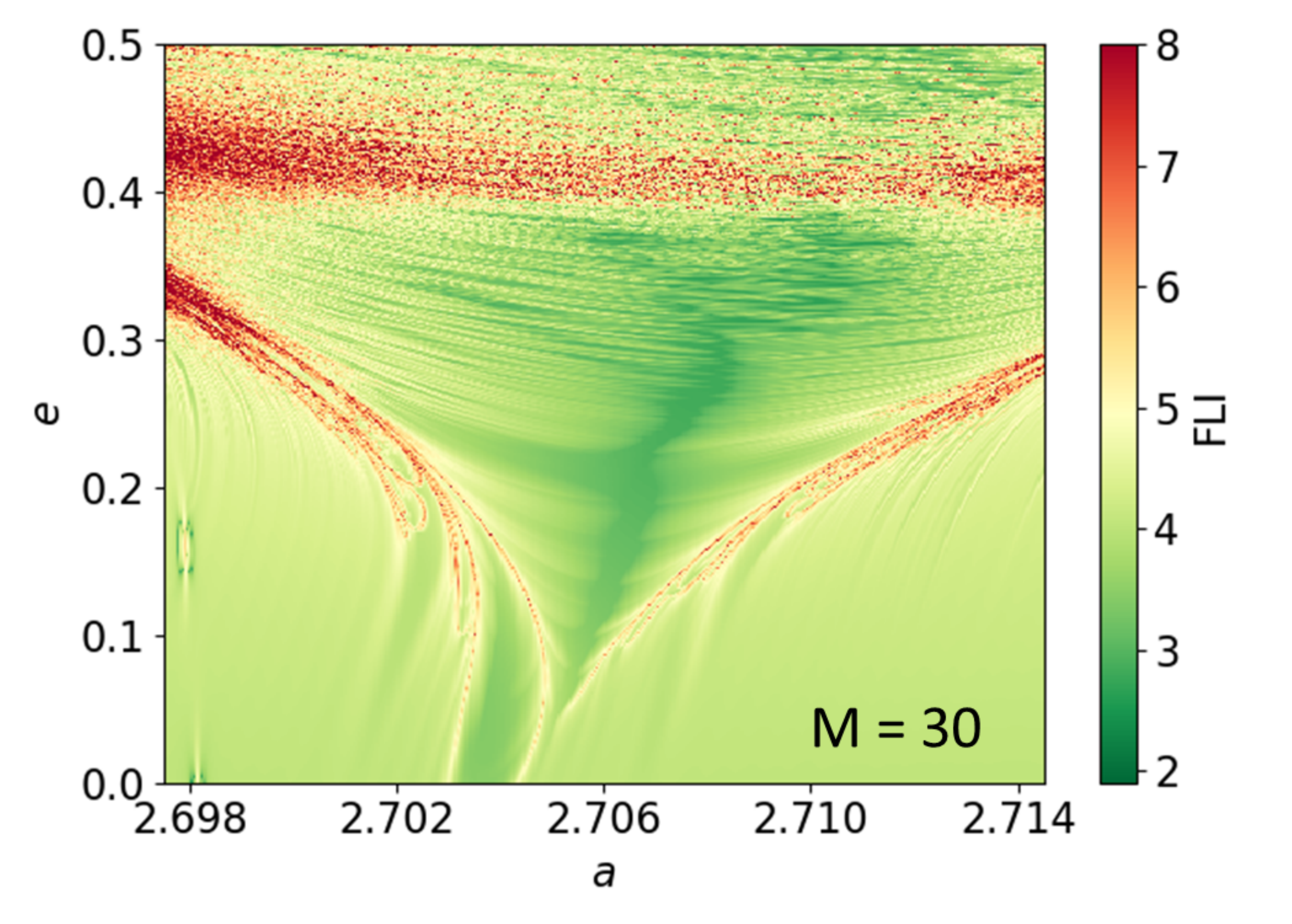}}
	\caption{FLI maps of the 8:3 MMR with Jupiter in orbital plane of Ceres ($i=10.594^\circ$, $\omega=73.37^\circ$, $\Omega=80.306^\circ$) for mean anomaly (a) $M=60^\circ$ and (b) $M=30^\circ$. The maps were computed for $5\,kyr$ on a grid of $400 \times 400$ initial conditions (test particles), whereas semi-major axis $a \in \left\langle 2.6975, 2.7145 \right\rangle \, AU$ and eccentricity $e \in \left\langle 0, 0.5 \right\rangle$. Colour scale represents FLI values determined as a maximum from values obtained for three different initial deviation vectors.}
	\label{400x400}
\end{center}
\end{figure*}

\begin{figure*}
\begin{center}
	\subfloat[]{\includegraphics[width=0.4\textwidth]{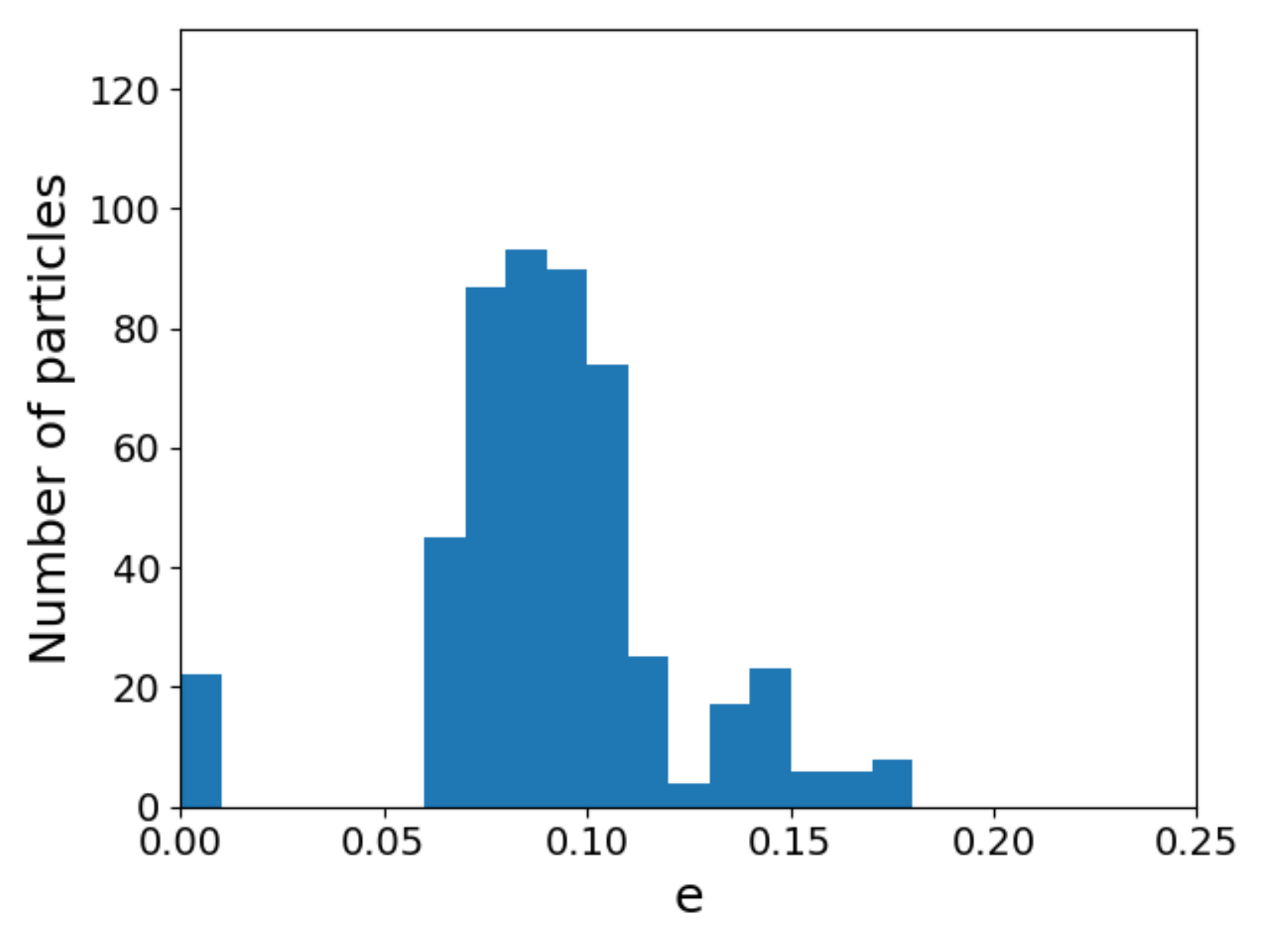}}
	\subfloat[]{\includegraphics[width=0.4\textwidth]{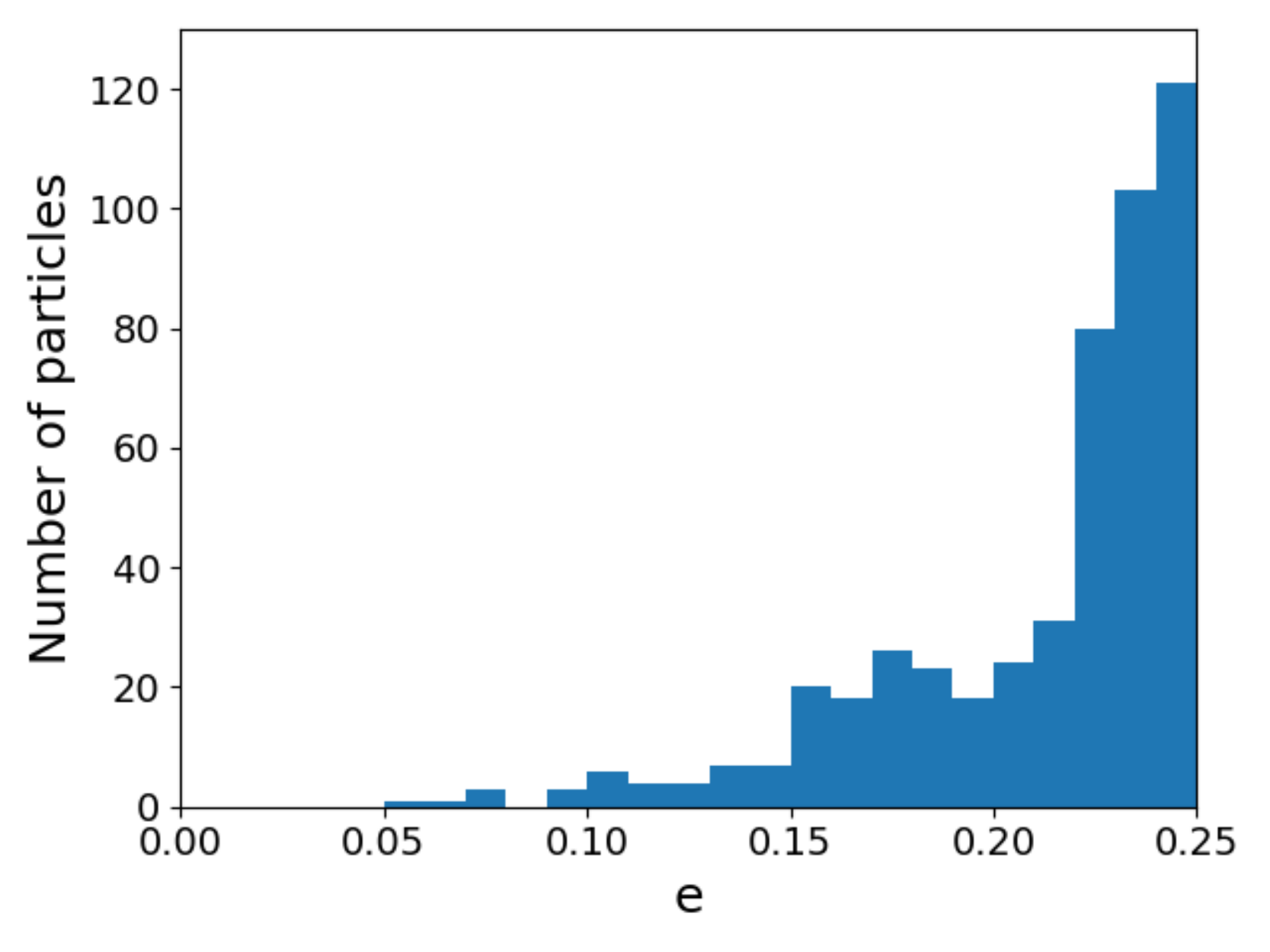}}
	\caption{Initial distribution of eccentricities for (a) 500 of the most stable particles and (b) 500 of the most unstable particles in the case of mean anomaly $M=60^\circ$.}
	\label{hist_e}
\end{center}
\end{figure*}

\begin{figure*}
\begin{center}
	\subfloat[]{\includegraphics[width=0.4\textwidth]{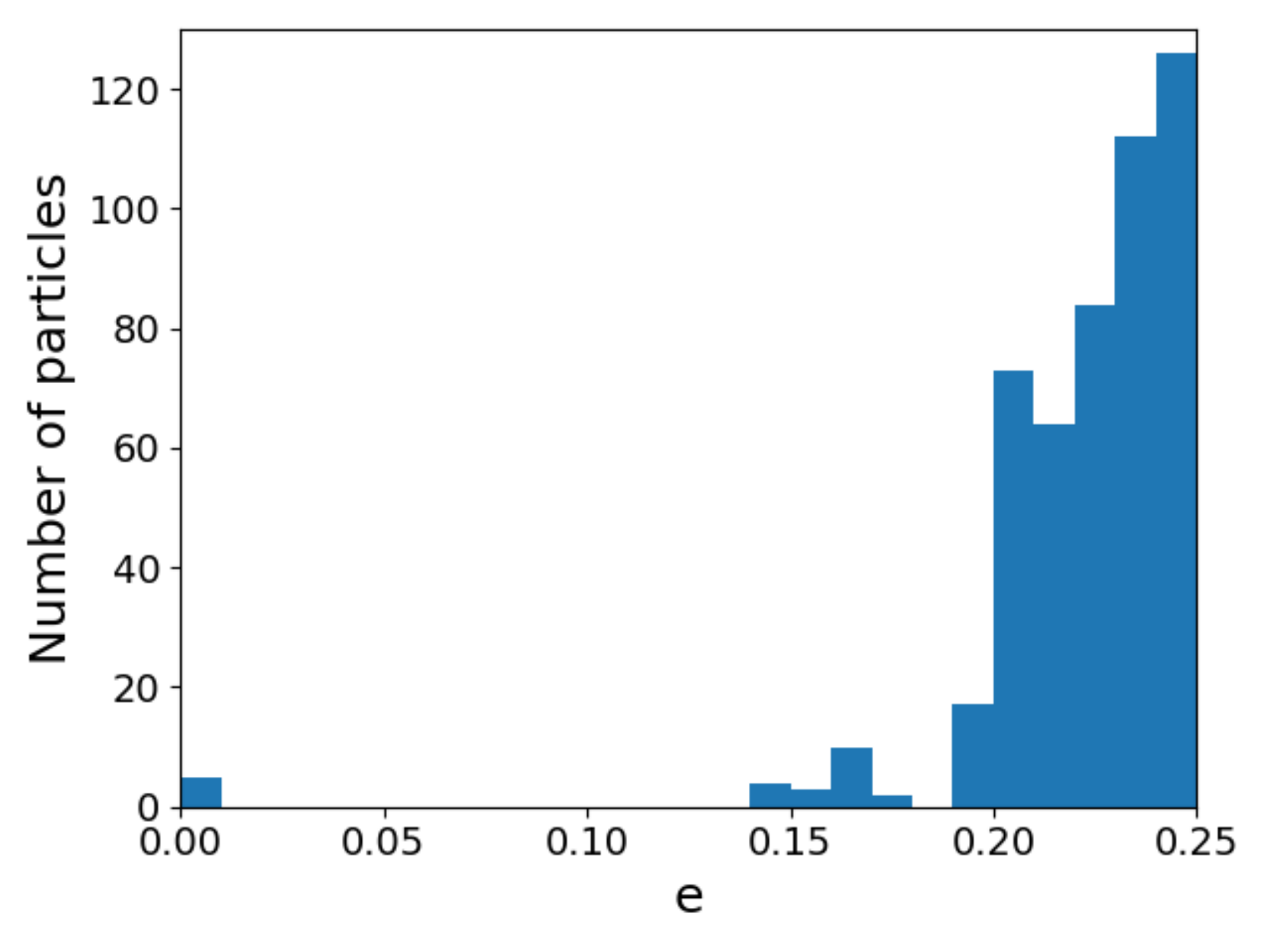}}
	\subfloat[]{\includegraphics[width=0.4\textwidth]{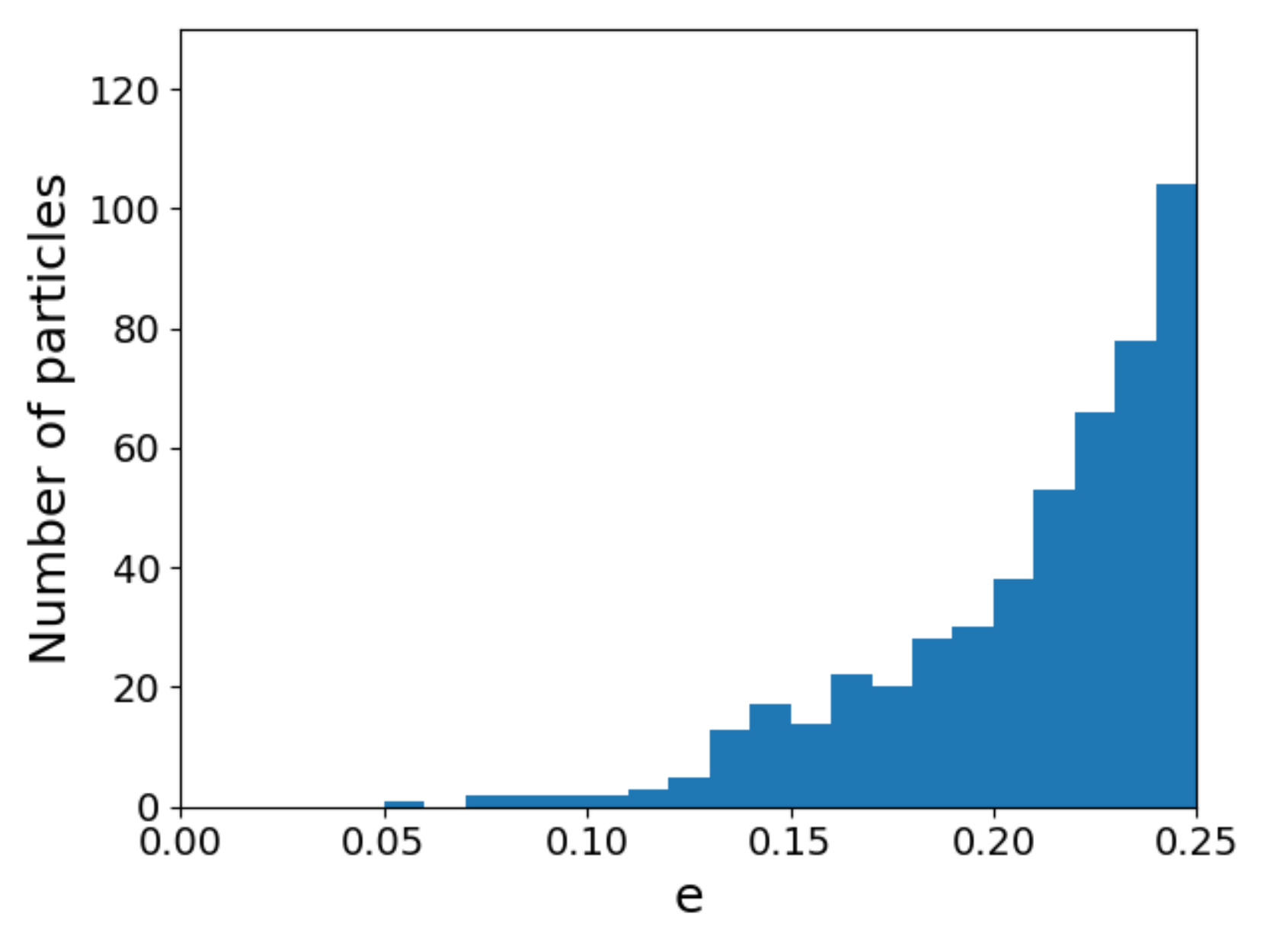}}
	\caption{Initial distribution of eccentricities for (a) 500 of the most stable particles and (b) 500 of the most unstable particles in the case of mean anomaly $M=30^\circ$.}
	\label{hist_eM30}
\end{center}
\end{figure*}

We decided to map the 8:3 MMR with Jupiter again, but this time on a grid of $400\times 400$ initial conditions (see Fig. \ref{400x400}). These maps were computed by the same procedure as  the maps on the Fig. \ref{50x50a}-\ref{50x50b}. This means we considered the Sun and all of the planets. The maps were integrated in the orbital plane of Ceres for $5000\,yr$ by WHFast integrator. The only modification (except the resolution) is that the colour scale represents the FLI value that was obtained as a maximum of 3 (not 6) different initial deviation vectors. This time, we tracked only the unit vectors corresponding to varying three components of the position. Our tests showed that, by reducing the number of initial deviation vectors, we obtained essentially the same maps, but it significantly shortened the integration time.

Based on the map for mean anomaly $M=60^\circ$ (Fig. \ref{400x400} (a)), we chose the 500 test particles with the highest ($9.8794\geq FLI \geq 6.0688$) and the 500 particles with the lowest ($2.6291 \leq FLI \leq 3.2364$) FLI values from the region with eccentricities $e\leq 0.25$. We limited the eccentricity because we do not expect material released from Ceres ($e=0.078$) to have high values of eccentricity. As you can see in Fig. \ref{hist_e}, most of the unstable particles already started with higher eccentricities in comparison to stable particles,  so their orbits were less circular at the beginning of the simulation. Analogously, we also chose particles based on the FLI map for mean anomaly $M=30^\circ$. Initial distributions of eccentricities in this case are shown in Fig. \ref{hist_eM30}.
Selected particles from both maps were integrated for $15\,Myr$ by the MERCURIUS integrator because our test simulations showed that the MERCURIUS is more reliable in the case of long-term integrations than the  WHFast integrator. For the sake of simplicity, we were not detecting collisions or close approaches during the simulation. For our purposes, we decided to use output step $100\,yr$. Integrations were stopped if the particle reached the hyperbolic orbit ($e>1$ or $a<0\,AU$) or an orbit with semi-major axis $a>100\,AU$. 

\section{Mean anomaly $M = 60^\circ$}
\subsection{500 of the most stable particles}

\begin{figure}
\centering
  \includegraphics[width=1\linewidth]{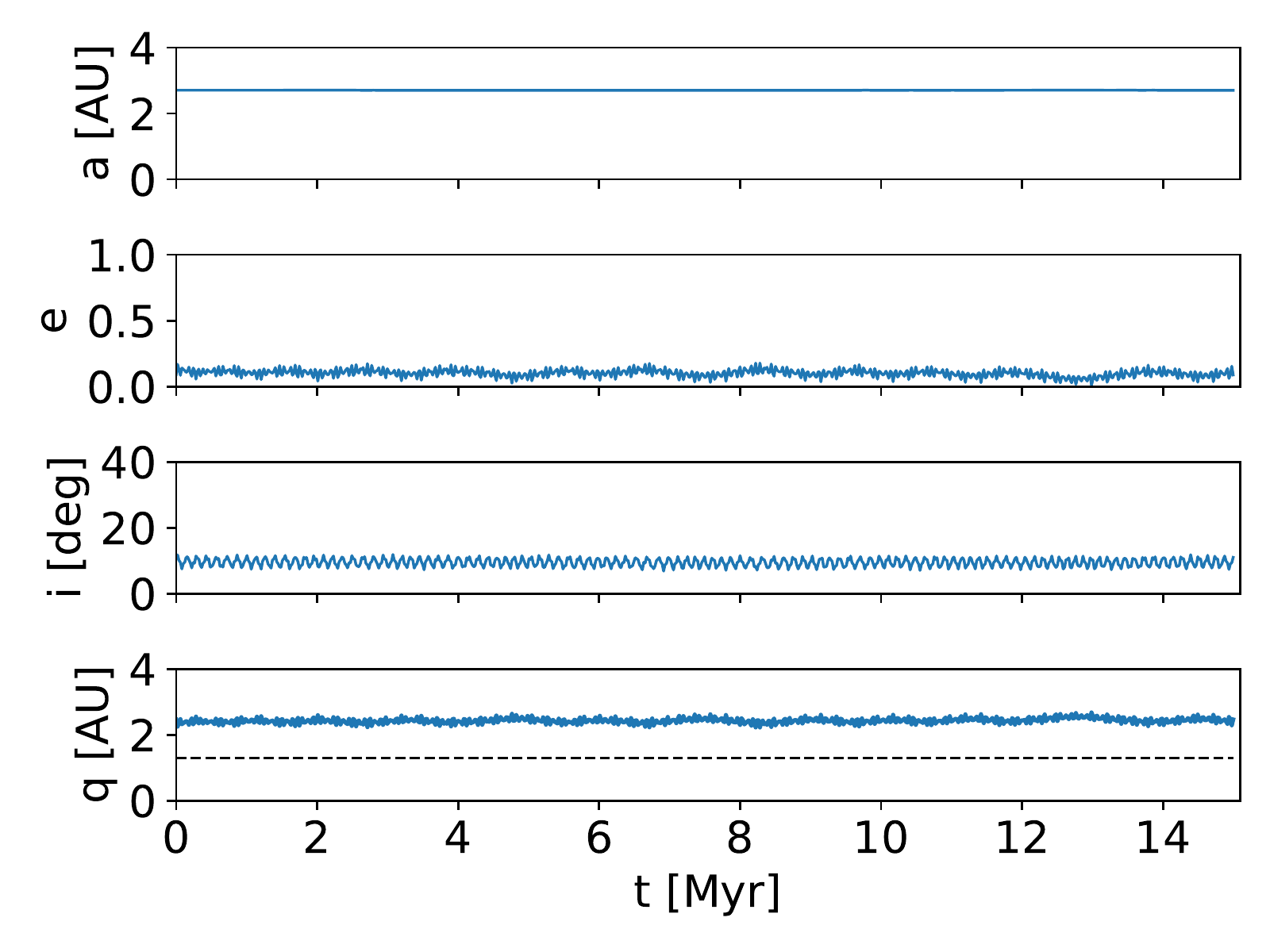}
\caption{The typical orbital evolution for a stable particle (a particle with a small FLI value) with $M=60^\circ$. Dashed black line inside the graph for perihelion distance represents $q=1.3\,AU$. }
\label{stable}
\end{figure}

\begin{figure*}
\begin{center}
	\subfloat[]{\includegraphics[width=0.4\textwidth]{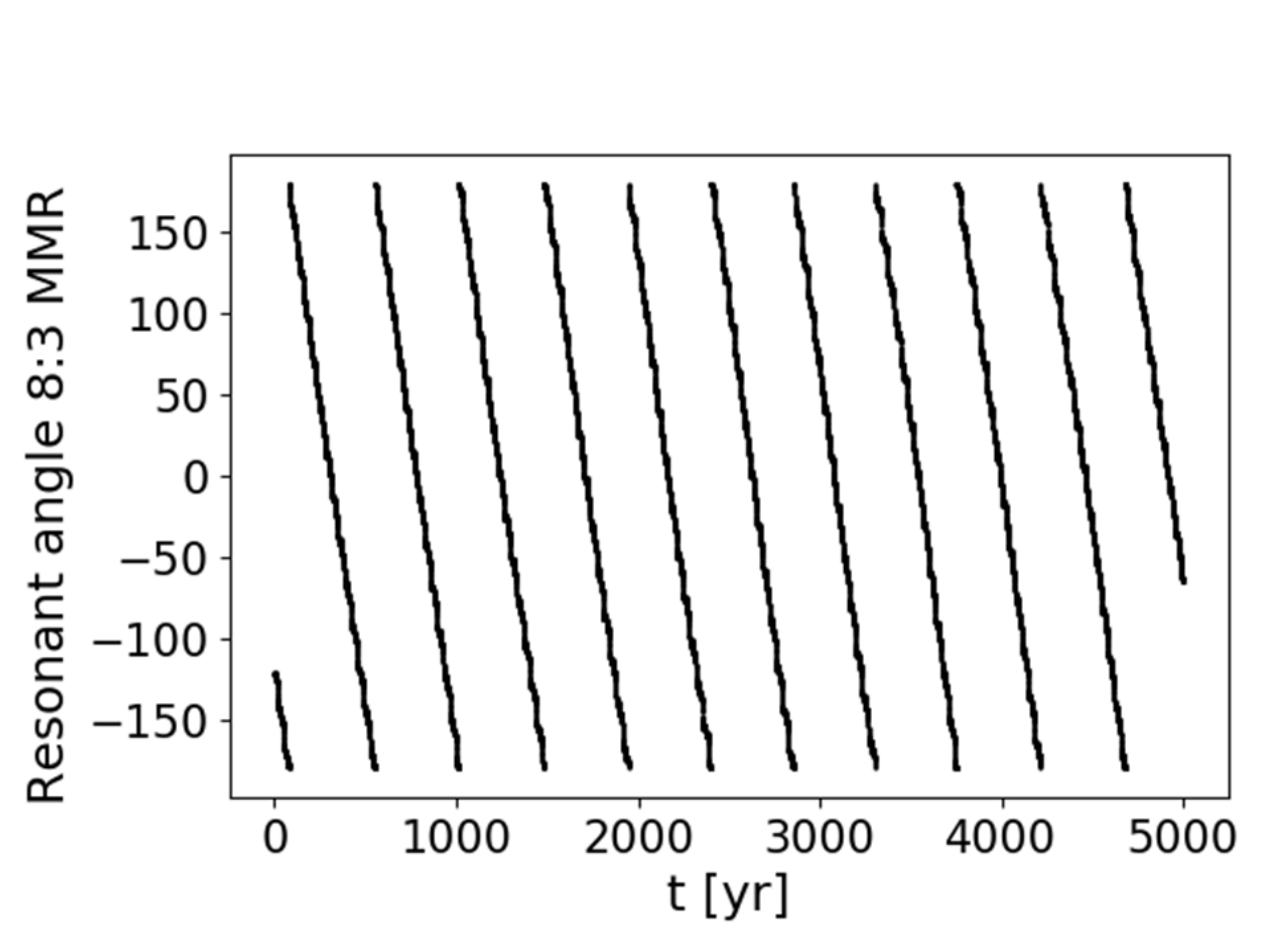}}
	\subfloat[]{\includegraphics[width=0.4\textwidth]{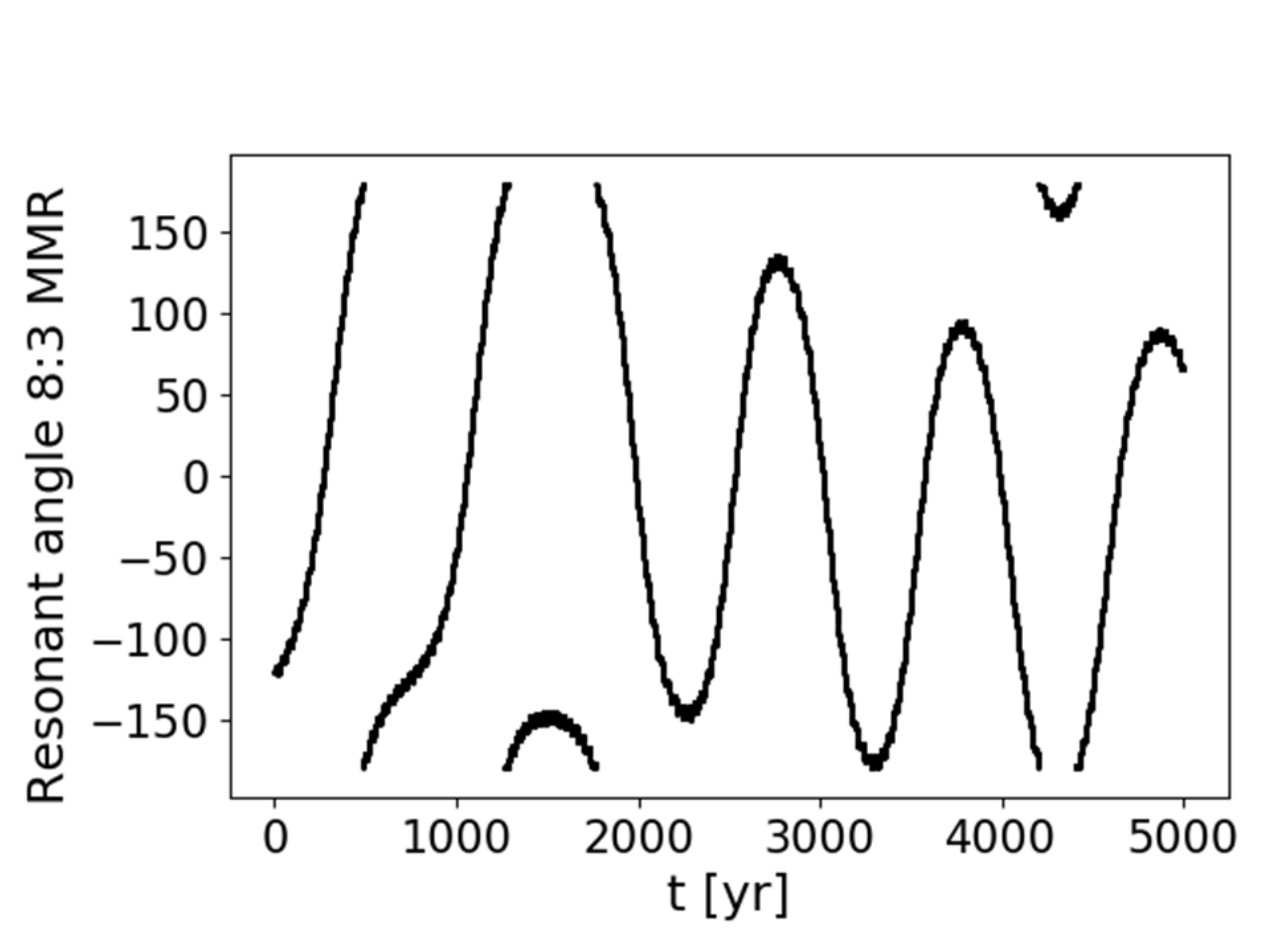}}
	\caption{The evolution of the resonant angle for (a) a stable and (b) an unstable particle ($M=60^\circ$). The resonant angle values are in degrees.}
	\label{resonant}
\end{center}
\end{figure*}

Orbital evolution of the most stable particles (particles with the lowest FLI values) was stable in general. The typical orbital evolution for a stable particle is shown in Fig. \ref{stable}. Simulations for only two particles were stopped because they managed to reach the orbit with $a>100\,AU$. Several other particles reached the orbit with a slightly increased eccentricity. During the simulation 38 ($7.6\,\%$) test particles reached the perihelion distance $q\leq 2\,AU$ at some point. Only 4 ($0.8\,\%$) particles managed to reach the limit of $q\leq 1.3\,AU$, which is necessary for a particle to be classified as a near-Earth object (NEO). The perihelion distance $q \leq 1\,AU$ was reached by only two particles for which the integration stopped before $15\,Myr$ because of excessively high values of semi-major axis. An additional short-term simulation of 30 of the most stable particles (for $5\,kyr$), with a more frequent output, showed that these stable particles did not show any signs of capturing in the 8:3 MMR with Jupiter. That means the resonant angle did not oscillate around a particular value (see Fig. \ref{resonant} (a)).

\begin{figure*}
\begin{center}
	\subfloat[]{\includegraphics[width=0.45\textwidth]{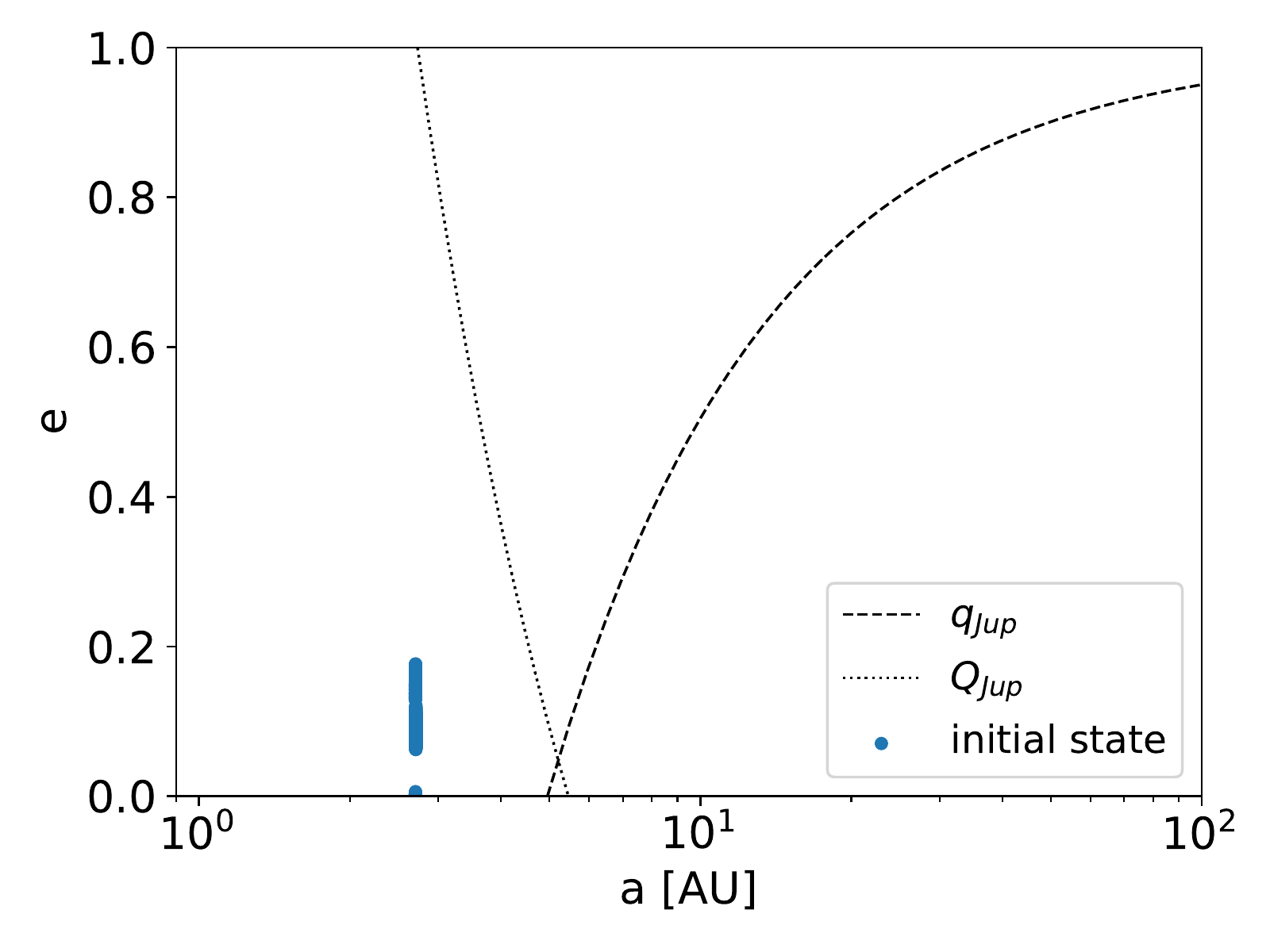}}
	\subfloat[]{\includegraphics[width=0.45\textwidth]{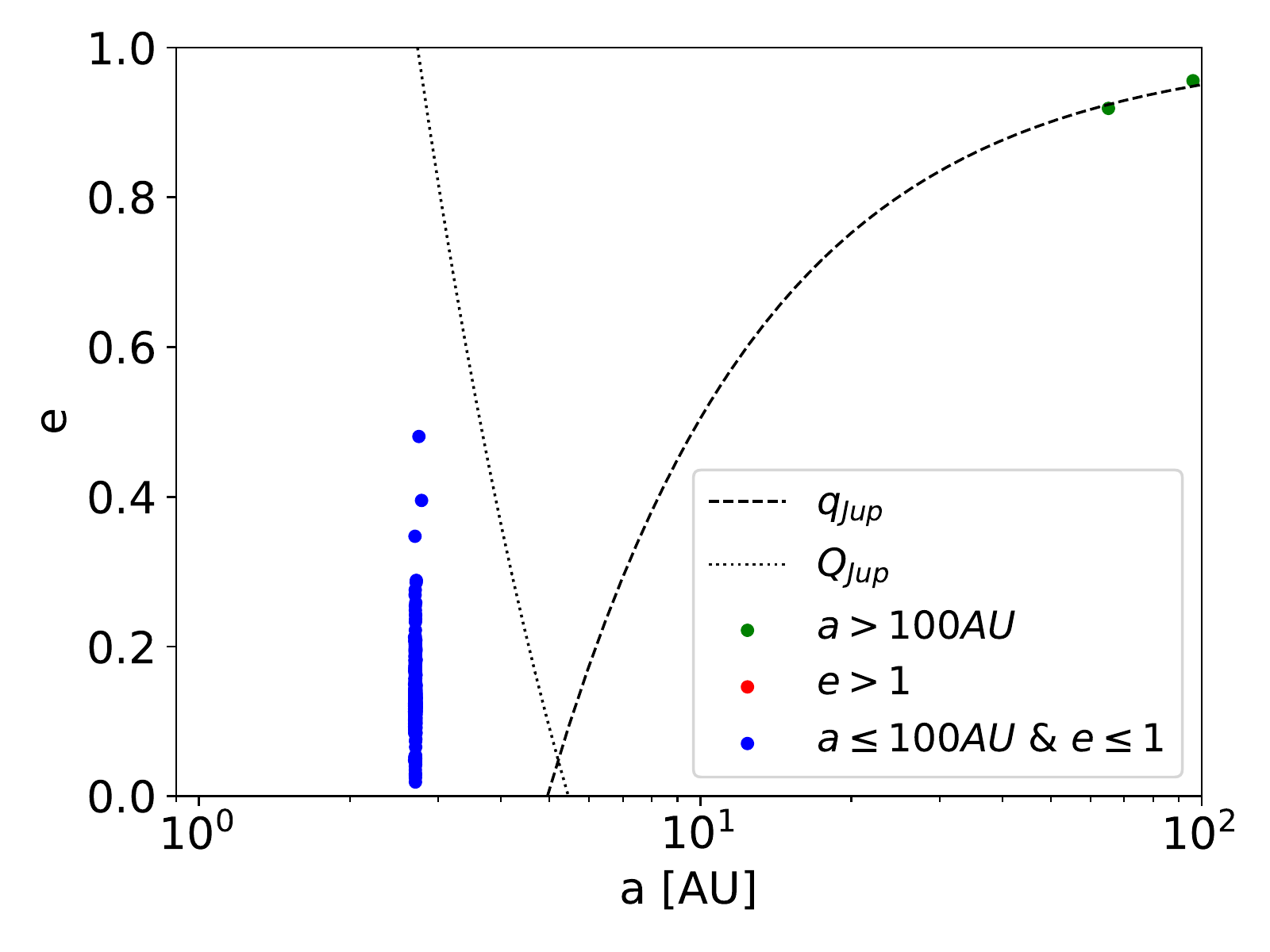}}
	\caption{Initial (a) and end (b) states of the simulation for stable particles ($M=60^\circ$). If particles reached an orbit with $a>100\,AU$ (green dots) or a hyperbolic orbit (red dots), their last known parameters before they were ejected to these orbits are displayed. Dashed black line represents the perihelion distance of Jupiter; dotted black line represents the aphelion distance of Jupiter.}
	\label{init_end_stable}
\end{center}
\end{figure*}

Fig. \ref{init_end_stable} represents the initial and final states of the simulation for stable particles. If the particle reached $a>100\,AU$, the figure shows the last known values that satisfied the condition $a\leq 100\,AU$. In the figure, there are also black lines that characterise the perihelion and aphelion distance of Jupiter. As you can see, the two particles that reached $a>100\,AU$ (green dots) are located along the line characterized by the perihelion distance of Jupiter. That means the perihelion distance of these test particles was close to the perihelion distance of Jupiter. So it can be assumed that Jupiter is responsible for the final state of these two particles.

\subsection{500 of the most unstable particles}
\begin{figure}
\centering
  \includegraphics[width=0.9\linewidth]{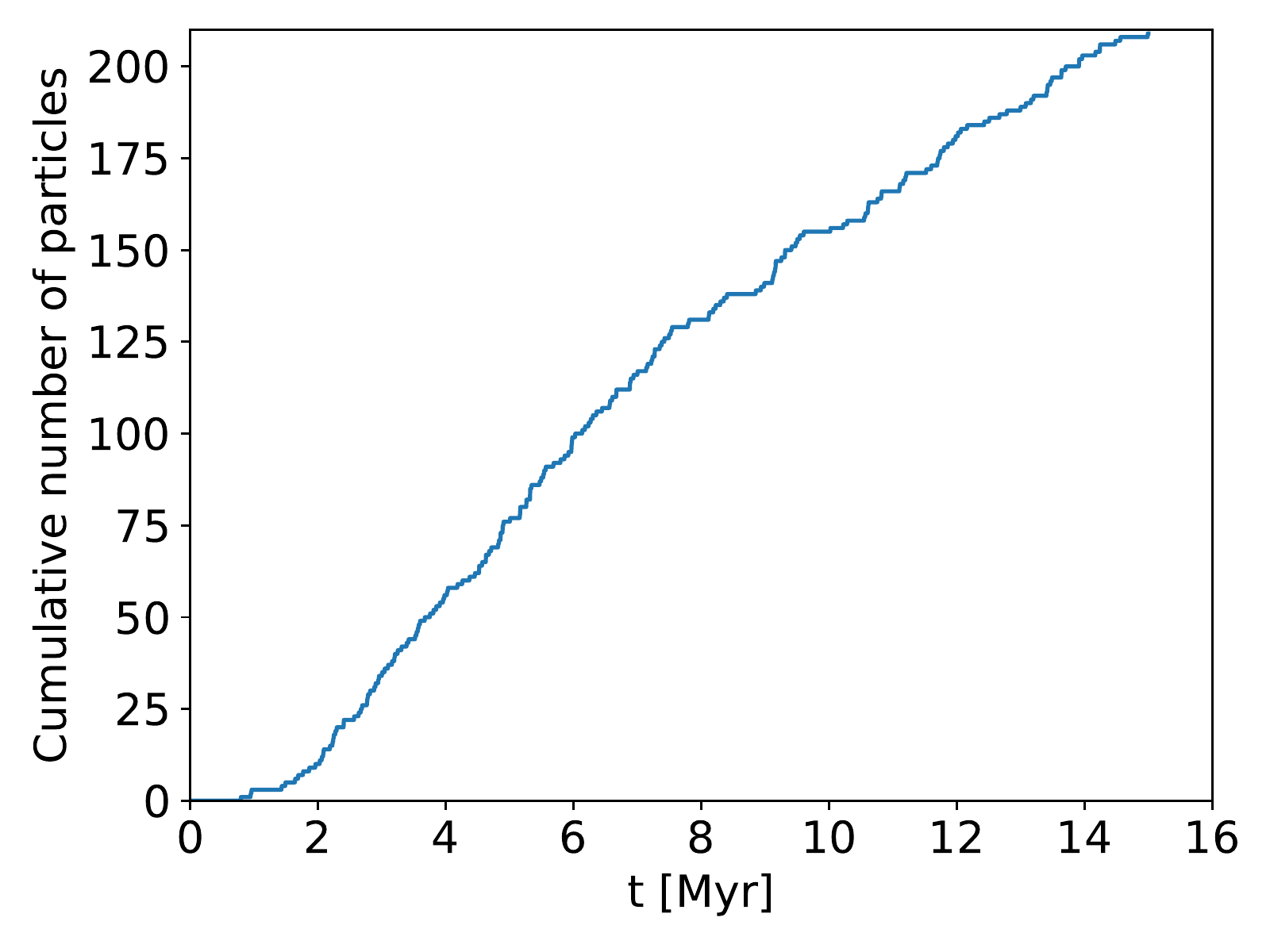}
\caption{Cumulative number of particles (from 500 of the most unstable particles) that reached $q \leq 1.3\,AU$.}
\label{kumulat}
\end{figure}

The orbital evolution of particles with the highest FLI values was significantly different from the evolution of stable particles. However, a few unstable particles also had an orbital evolution similar to the typical evolution of stable particles (Fig. \ref{stable}). Simulations for 138 ($27.6\,\%$) particles were stopped before finishing $15\,Myr$. In 97 cases, this was because the particle reached the orbit with $a>100\,AU$. The remaining 41 particles reached a hyperbolic orbit with eccentricity $e>1$. Out of the 500 particles, 469 ($93.8\,\%$) reached the orbit with a perihelion distance $q\leq 2\,AU$ at some point during the simulation. In 209 ($41.8\,\%$) cases, the limit of $q\leq 1.3\,AU$ was reached. 177 particles reached the orbit with $q\leq 1\,AU$, which means that these particles could theoretically be on Earth-crossing orbits. Fig. \ref{kumulat} shows the cumulative number of particles that managed to reach the limit of $q\leq 1.3\,AU$. As can be seen, the first particles with $q\leq 1.3\,AU$ started to appear after approximately $1\,Myr$.

\citet{sungrazer} define sungrazing object in case of $q<3.45R_\odot\approx0.016\,AU$ and so-called sundiving object (or Sun-impactor) when $q<R_\odot\approx 0.00465\,AU$ (where $R_\odot$ represents the radius of the Sun). During simulation, 46 ($9.2\,\%$) particles reached the orbit with $q<0.016\,AU$ and 41 of them ($8.2\,\%$ of 500) were found on a sundiving orbit at some point. All of these particles reached this type of orbit by increasing their eccentricity close to the value of 1 and not by any significant decrease in semi-major axis. Since our simulations were not detecting collisions or close approaches, we can only suppose that at least several of these sungrazing particles would finish their evolution by impacting the Sun. We remark that integrations, after a possible close approach to the Sun, are very uncertain. 
For example, according to our simulations, five sungrazing particles that did not become sundivers ended up on an orbit with $a>100\,AU$.
 \citet{gladm} registered that $17.2\,\%$ particles with an origin in 8:3 MMR with Jupiter ended their evolution by impacting the Sun. On the other hand, they discovered that $35.7\,\%$ particles ended up behind the orbit of Saturn. In our simulations, we registered 142 ($28.4\%$) particles behind Saturn. We should note that the conclusions of \citet{gladm} are based on a simulation of 157 particles for the Chloris family (mean orbital elements: $a=2.746\,AU$, $e=0.252$, $i=8.783^\circ$ \citep{families}) for $29\,Myr$.

\begin{figure}
\centering
  \includegraphics[width=\linewidth]{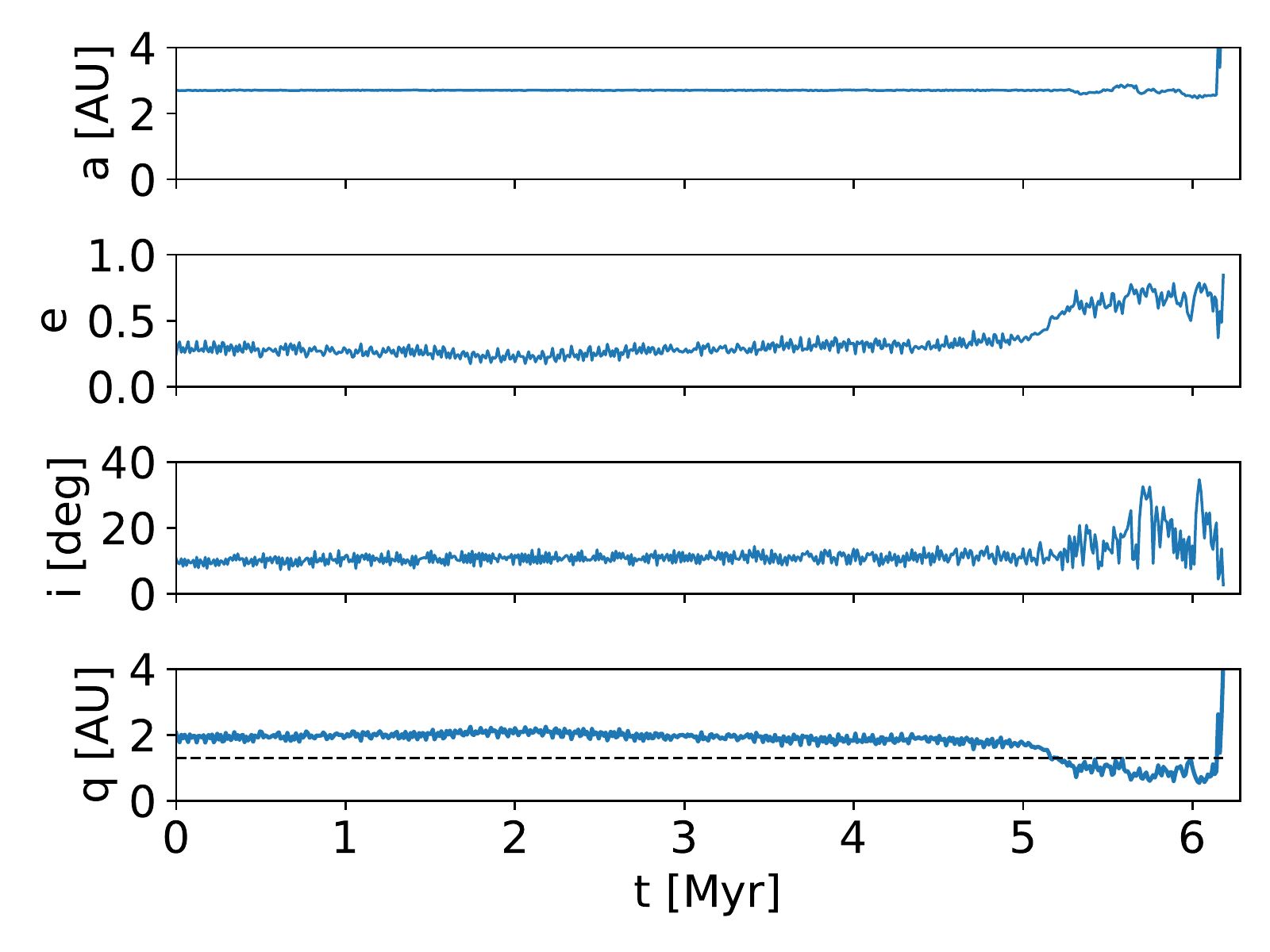}
\caption{The orbital evolution for an unstable particle (a particle with a large FLI value) with $M=60^\circ$. Dashed black line inside the graph for perihelion distance represents $q=1.3\,AU$.}
\label{unstable}
\end{figure}

In this case, an additional short simulation of 30 of the most unstable particles (for $5\,kyr$) revealed that these particles show signs of capturing in 8:3 MMR with Jupiter, when the resonant argument oscillates (Fig. \ref{resonant} (b)). We also performed an integration with a more frequent output (every $0.25\,yr$) for 10 particles, which did not finish the whole integration for $15\,Myr$, to find out what happened to them. First of all, these particles experienced an increase in eccentricity while their semi-major axis remained approximately the same. That means, the particles were getting closer to inner planets. Subsequently, perturbations of the inner planets caused the particles to eventually reach Jupiter, which was responsible for the final state of these particles, i.e., $a>100\,AU$ or $e>1$. 
 As an example, we show the orbital evolution for an unstable particle (Fig. \ref{unstable}) which took approximately $6\,Myr$. This particular particle reached the hyperbolic orbit.

\begin{figure*}
\begin{center}
	\subfloat[]{\includegraphics[width=0.5\textwidth]{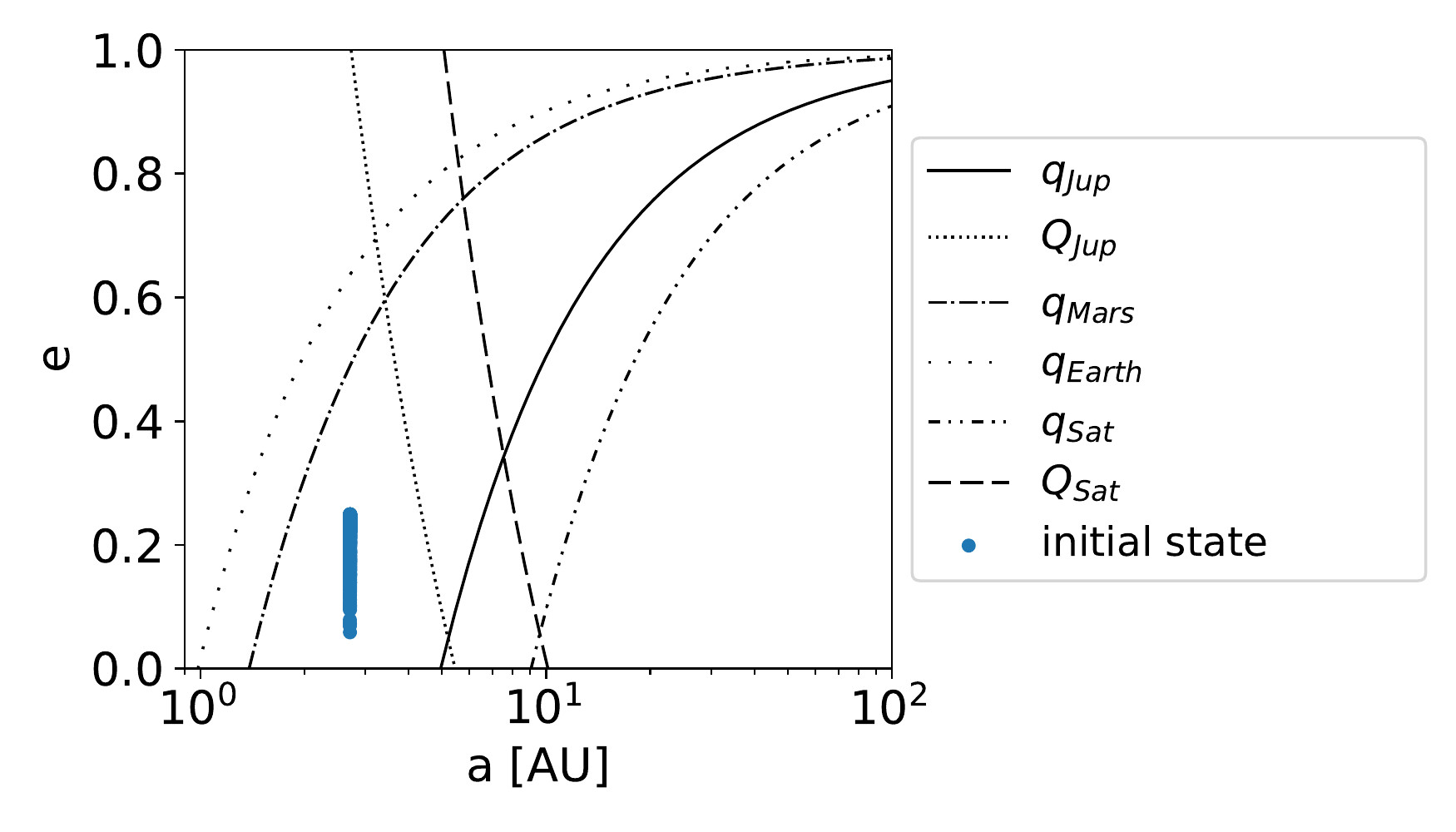}}
	\subfloat[]{\includegraphics[width=0.5\textwidth]{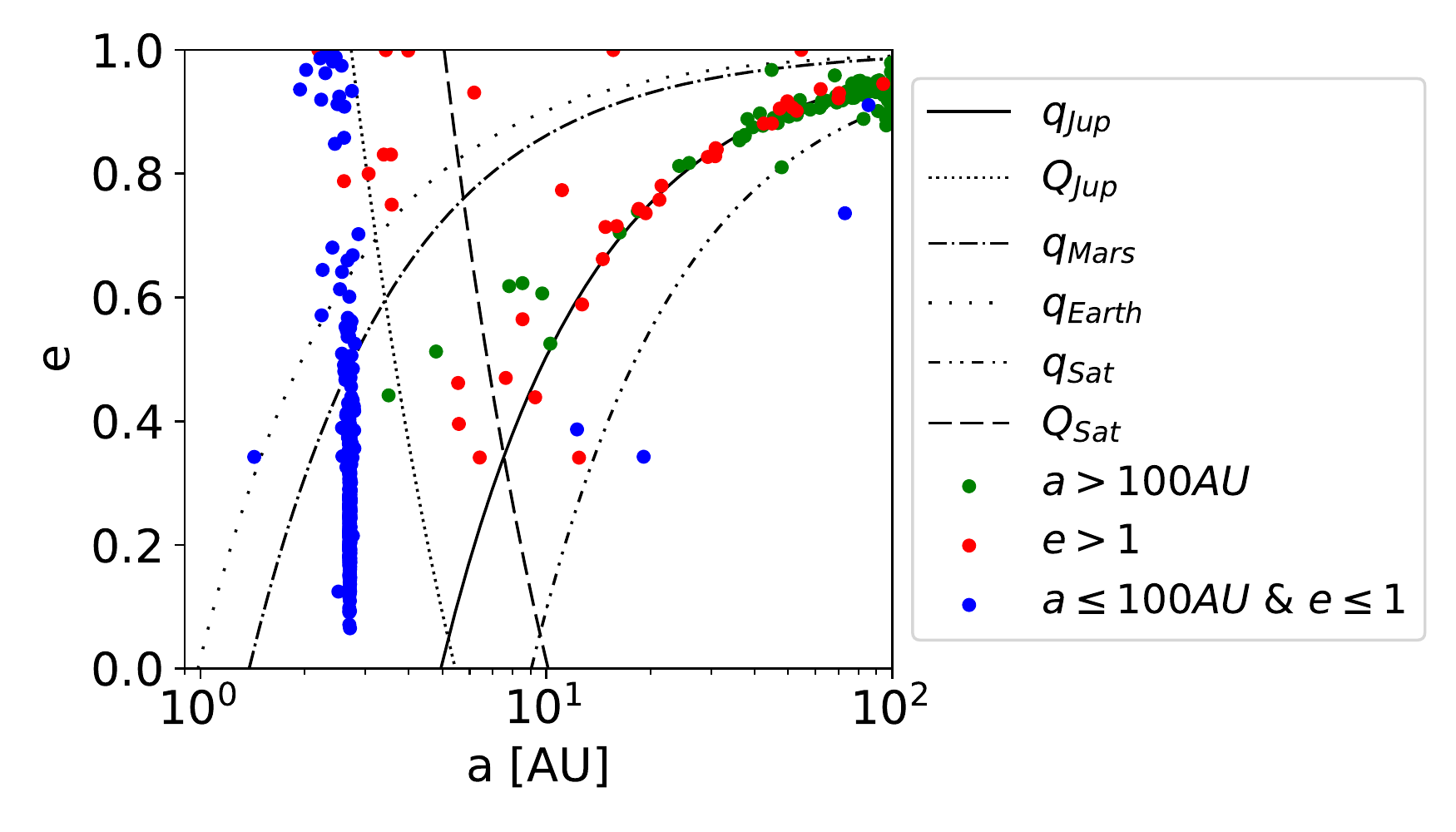}}
	\caption{Initial (a) and end (b) states of the simulation for unstable particles ($M=60^\circ$). If particles reached an orbit with $a>100\,AU$ (green dots) or a hyperbolic orbit (red dots), their last known parameters before they were ejected to these orbits are displayed. Black lines represent perihelion and aphelion distances of selected planets.}
	\label{init_end_unstable}
\end{center}
\end{figure*}

Initial and end states of the simulation for unstable particles are presented in Fig. \ref{init_end_unstable}.  
The red and green dots in the figure represent the positions from which the particles were ejected to the hyperbolic orbit or to an orbit with $a>100\,AU$.
As you can see in Fig. \ref{init_end_unstable}, most of the red and green dots are placed along, or in-between, the lines representing the perihelion and aphelion distances of Jupiter. So we assumed that Jupiter plays a dominant role in the final state of unstable particles. However, other planets such as Earth, Mars or Saturn also contribute to the scattering of test particles.

The 8:3 MMR with Jupiter is not usually treated as a potential source of NEOs. The works dealing with modelling of the NEO population, such as \citet{bottke} or \citet{granvik}, do not consider this resonance as a potential source region or escape route. Our simulations suggest that it is possible for 8:3 MMR to transport material to the NEO region or close to the Earth, potentially even material from Ceres.

This 8:3 MMR has been studied in combination with 5:2 MMR by \citet{todor} and \citet{deleon}. The authors investigated the link between Pallas (or the Pallas family) and Phaethon. \citet{todor} using the FLI map in the orbital plane of Pallas (for which $a=2.734\,AU$, $e=0.230$, $i=34.854^\circ$) integrated 1000 particles for $5Myr$ in order to find particles on an orbit similar to the current orbit of Phaethon ($a=1.271\,AU$, $e=0.890$, $i=22.257^\circ$). $46.9\,\%$ of particles reached the desired type of orbit. Hence, \citet{todor} assumed that 5:2 and 8:3 MMRs with Jupiter possess a very powerful mechanism for transporting objects to the NEO region close to the asteroid Phaethon. During our simulations in the orbital plane of Ceres ($a=2.766\,AU$, $e=0.078$, $i=10.594^\circ$), we did not register any particles on an orbit similar to the orbit of Phaethon. \citet{deleon} integrated 1000 particles (for both resonances together) for $100\,Myr$. Of all these particles, 21 ($2.1\,\%$) spent some time on Mars- and Earth-crossing orbits with, $a<2\,AU$, before ending their lives. These results were interpreted as indicating that there exists a dynamical pathway, through which Pallas fragments may evolve into Phaethon-like orbits. In our study, we do not consider orbits with $a<2\,AU$ as important, but for completeness, during our simulation, we registered 15 ($3\,\%$) particles on an orbit with $a<2\,AU$.

\section{Mean anomaly $M=30^\circ$}
\subsection{500 of the most stable particles}

\begin{figure}
\centering
  \includegraphics[width=0.9\linewidth]{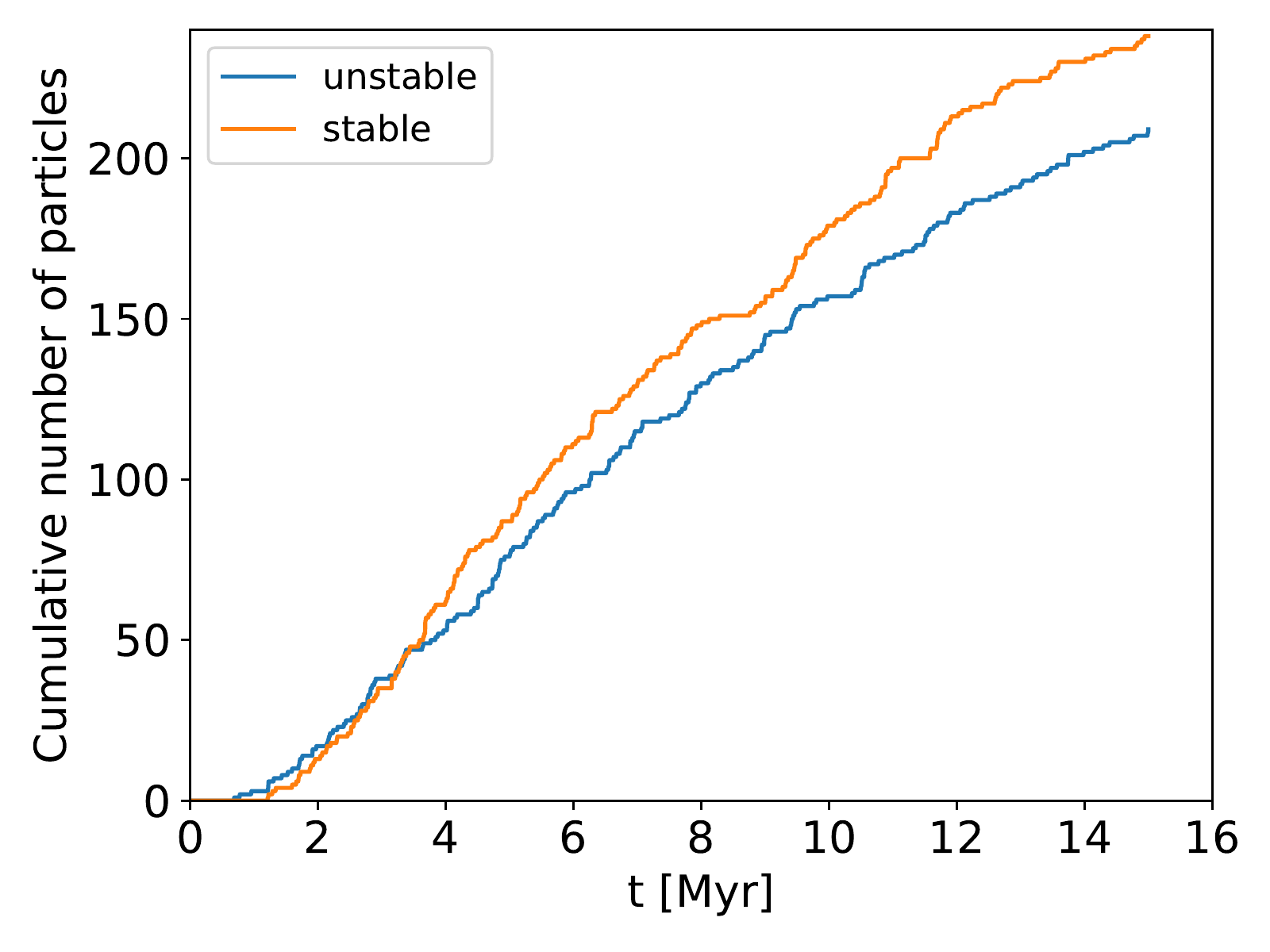}
\caption{Cumulative number of particles that reached $q \leq 1.3\,AU$ in the case of mean anomaly $M=30^\circ$. Orange line represents stable particles; blue line represents unstable particles.}
\label{kumulativeM30}
\end{figure}

\begin{figure*}
\begin{center}
	\subfloat[]{\includegraphics[width=0.4\textwidth]{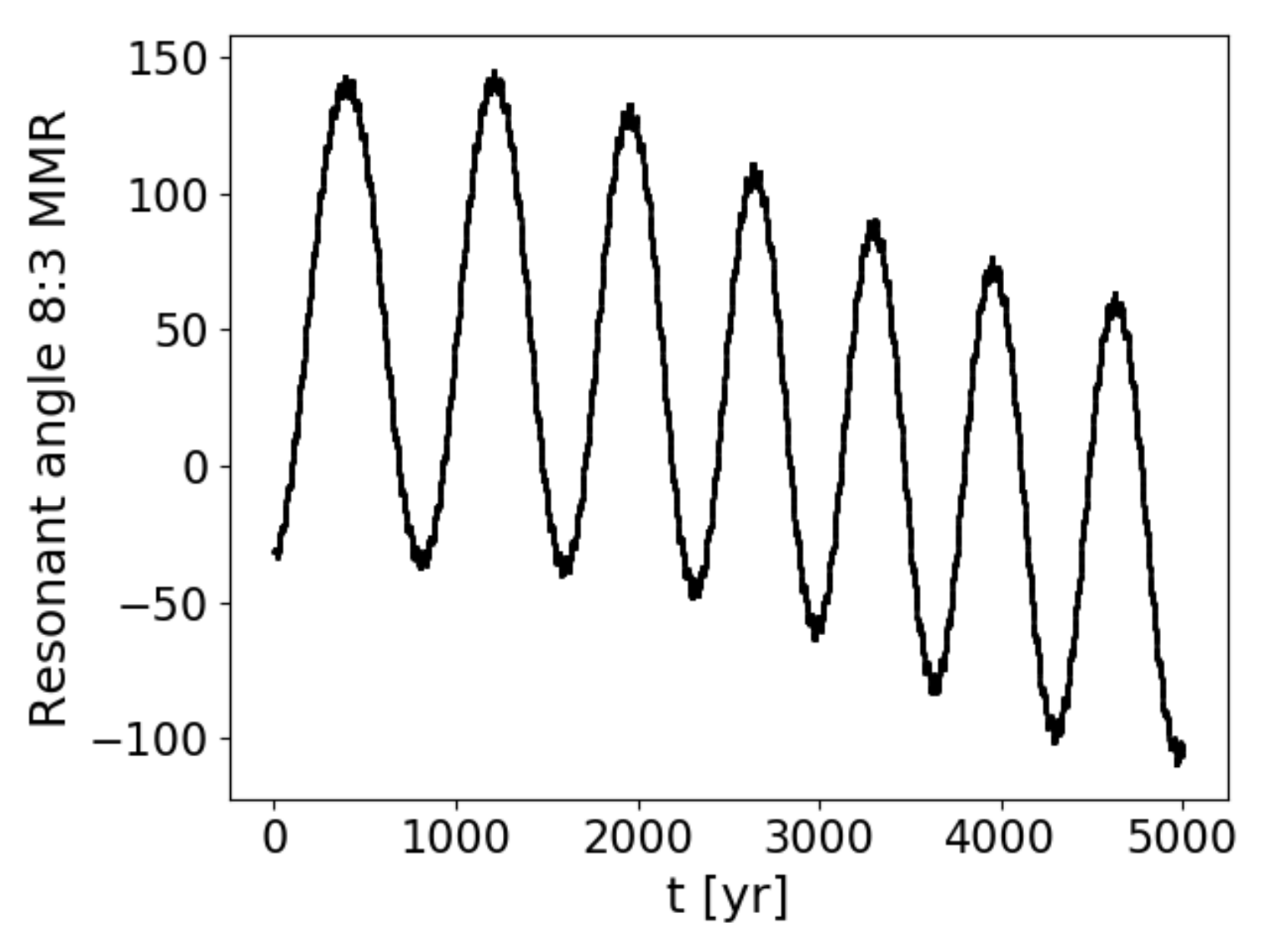}}
	\subfloat[]{\includegraphics[width=0.4\textwidth]{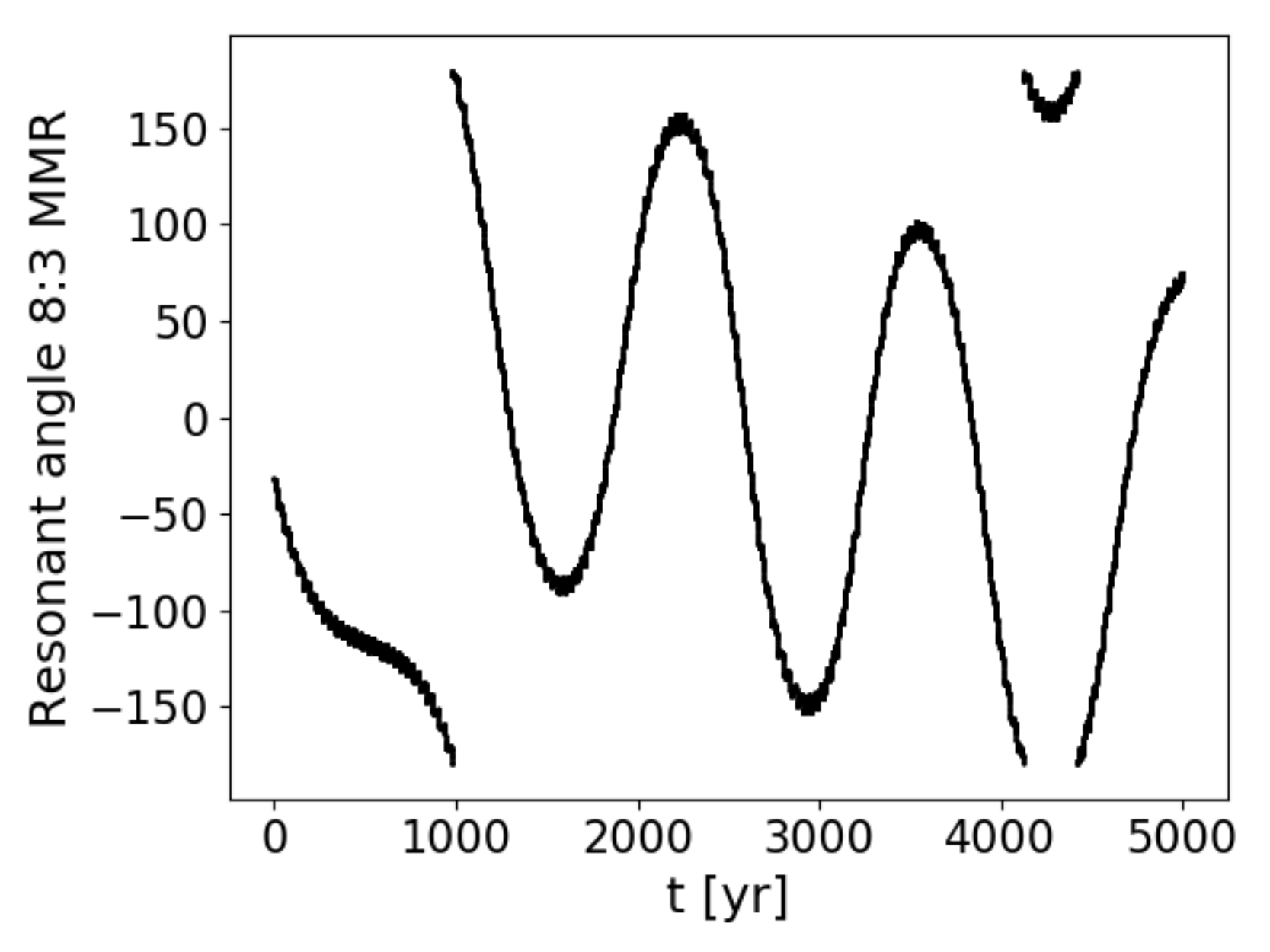}}
	\caption{The evolution of  the resonant angle for (a) a stable and (b) an unstable particle ($M=30^\circ$). The resonant angle values are in degrees.}
	\label{rezonM30}
\end{center}
\end{figure*}

\begin{figure*}
\begin{center}
	\subfloat[]{\includegraphics[width=0.5\textwidth]{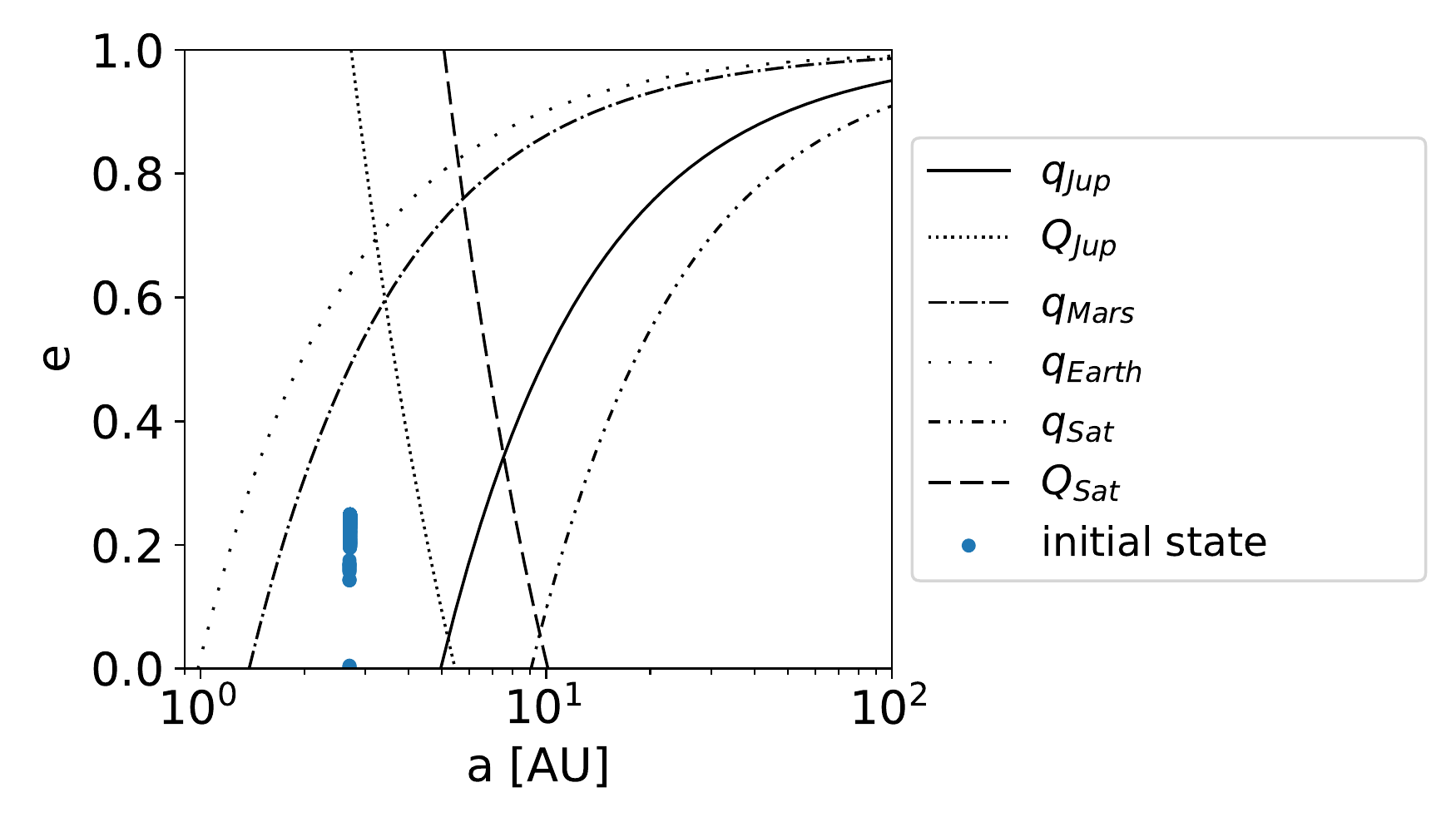}}
	\subfloat[]{\includegraphics[width=0.5\textwidth]{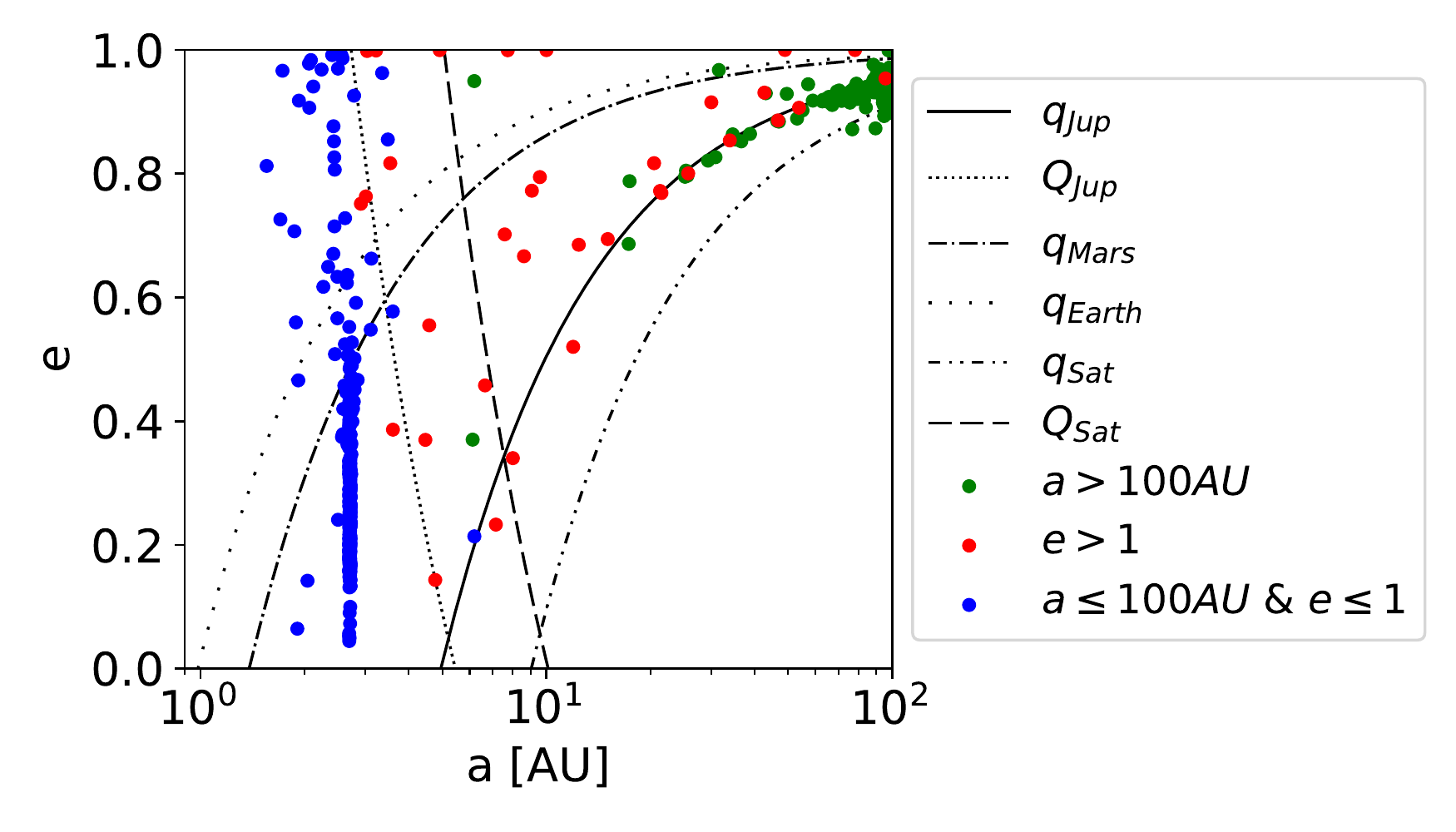}}
	\caption{Initial (a) and end (b) states of the simulation for stable particles ($M=30^\circ$). If particles reached an orbit with $a>100\,AU$ (green dots) or a hyperbolic orbit (red dots), their last known parameters before they were ejected to these orbits are displayed. Black lines represent the perihelion and aphelion distances of selected planets.}
	\label{init_end_greenM30}
\end{center}
\end{figure*}

Surprisingly, in the case of mean anomaly $M=30^\circ$, the orbital evolution of particles that were marked as stable, based on the FLI map, was not stable at all. During the simulation, 35 ($7\,\%$) particles were ejected to hyperbolic orbit and another 124 ($24.8\,\%$) particles reached an orbit with $a>100\,AU$. Out of 500 particles, 491 ($98.2\,\%$) reached the limit for perihelion distance $q\leq 2\,AU$ at some point during the simulation. 238 ($47.6\,\%$) particles reached the orbit with $q\leq 1.3\,AU$, and 202 of them ($40.4\,\%$ of 500) even reached $q \leq 1\,AU$. The cumulative number of particles that were able to reach the limit $q\leq 1.3\,AU$ is shown in Fig.~\ref{kumulativeM30}. These particles started to appear slightly after $1\,Myr$. This time, we registered 73 ($14.6\,\%$) sungrazing objects and from these 70 reached sundiving orbit at some point during the simulation. Additional short simulations of 30 of the most stable particles showed that most of these particles were captured in 8:3 MMR with Jupiter - the resonant angle oscillated (Fig.~\ref{rezonM30}). Similarly, as in the case of $M=60^\circ$, Fig. \ref{init_end_greenM30} represents the initial and final states of the simulation for stable particles. As can be seen, Jupiter again plays an important role in the orbital evolution of particles. This time, we registered 21 ($4.2\,\%$) particles on the orbit with $a<2\,AU$, and again we did not register any particles on an orbit similar to the current orbit of Phaethon.

The unexpected results for stable particles have several possible  explanations. The results may be affected by the fact that the majority of the stable particles started with a higher eccentricity ($e>0.2$) compared to the unstable particles, see Fig. \ref{hist_eM30}. In general, particles with a higher eccentricity drift away faster. Moreover, our FLI maps reflect the stability properties only for $5\,kyr$. It is not unusual that orbits become chaotic after $5\,kyr$ or after a million years or even later. For example, the particle in Fig. \ref{unstable} undergoes a significant change of orbital elements after approximately $5\,Myr$. In this case, we also plotted the FLI map computed for $15\,kyr$, to reveal the change in chaos. The FLI map did change but not dramatically. It may be supposed that the FLI maps captured "stable chaos", which has been well discussed ever since the work of \citet{stable-chaos}.  
The best numerical tool to detect stability is the Lyapunov characteristic exponents (LCE), which are calculated for extremely long time periods (in theory, infinite times). Short-term FLI maps are not reliable when searching for stable orbits, but they are very efficient in capturing the chaotic particles capable of fast transport.

\subsection{500 of the most unstable particles}

\begin{figure*}
\begin{center}
	\subfloat[]{\includegraphics[width=0.5\textwidth]{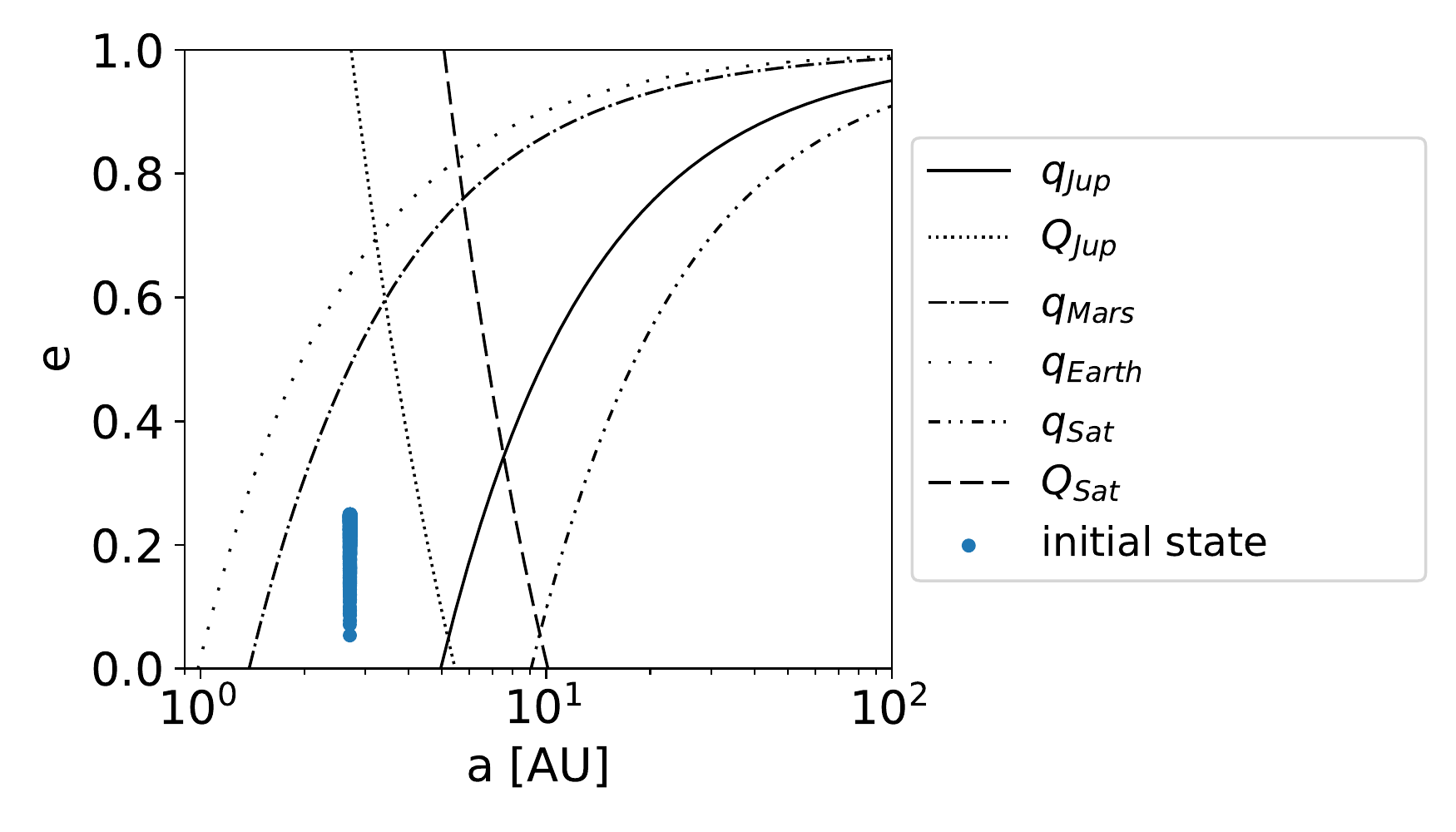}}
	\subfloat[]{\includegraphics[width=0.5\textwidth]{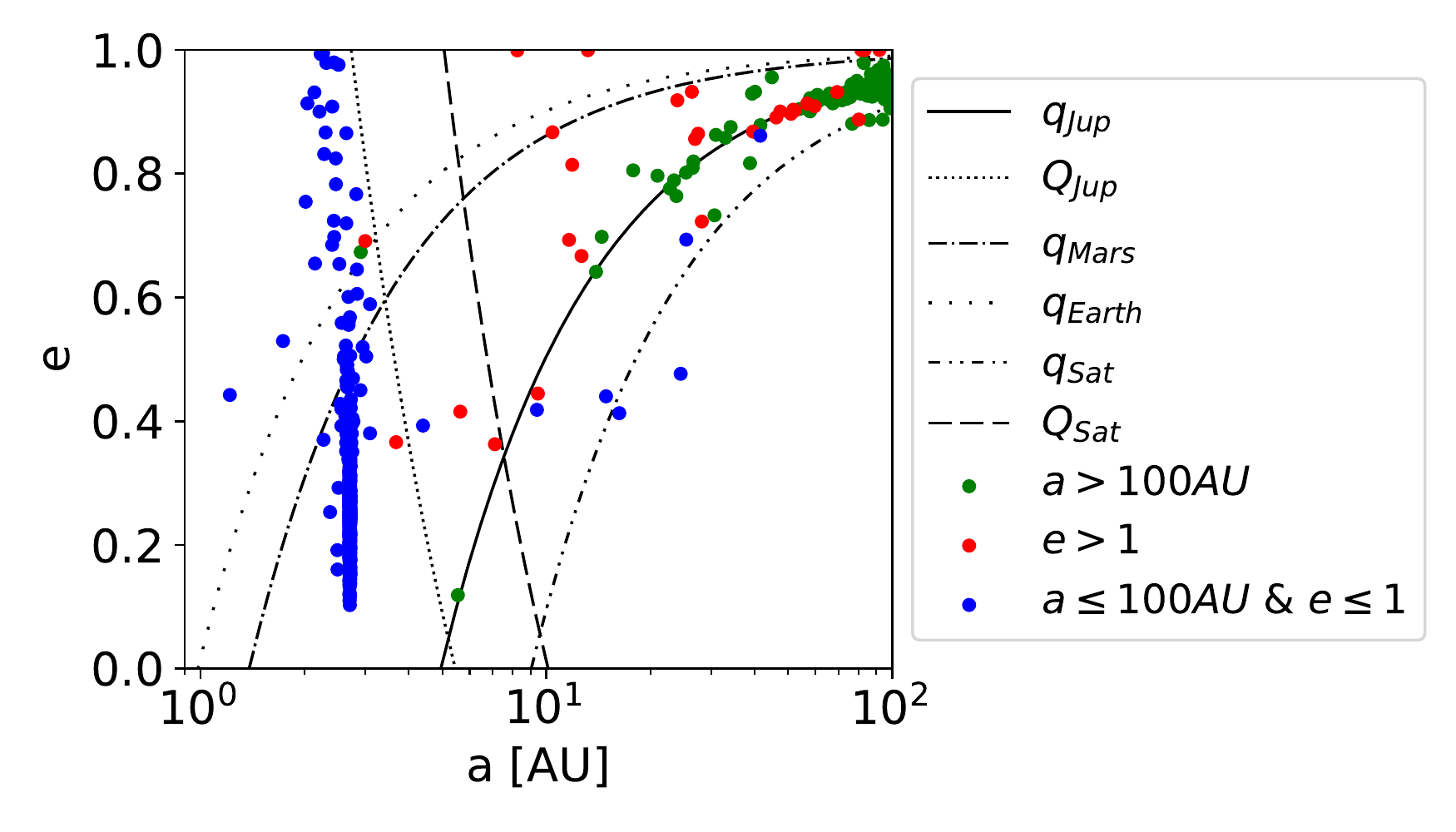}}
	\caption{Initial (a) and end (b) states of the simulation for unstable particles ($M=30^\circ$). If particles reached an orbit with $a>100\,AU$ (green dots) or a hyperbolic orbit (red dots), their last known parameters before they were ejected to these orbits are displayed. Black lines represent the perihelion and aphelion distances of selected planets.}
	\label{init_end_redM30}
\end{center}
\end{figure*}

In the case of unstable particles for mean anomaly $M=30^\circ$, simulations for 126 particles were stopped before finishing $15\,Myr$. In 97 ($19.4\,\%$ from 500) cases, it was caused by reaching the orbit with $a>100\,AU$. The remaining 29 ($5.8\,\%$ from 500) particles became hyperbolic. At some point during the simulation, 469 ($93.8\,\%$) particles reached $q\leq 2\,AU$, 209 of them ($41.8\,\%$ from 500) reached the limit $q \leq 1.3\,AU$, and 177 ($35.4\,\%$) were able to reach $q\leq 1\,AU$. Cumulative number of particles that were able to reach $q\leq 1.3\,AU$ is shown in Fig.~\ref{kumulativeM30}. In the case of unstable particles, we detected 50 ($10\,\%$) sungrazing particles, 44 ($8.8\,\%$) particles even reached a sundiving orbit. Short simulations for 30 of the most unstable particles revealed the resonant angle for almost all of the particles librates (Fig.~\ref{rezonM30}). This means these particles showed signs of capturing in 8:3 MMR with Jupiter. In Fig.~\ref{init_end_redM30}, you can see the initial and end states of the simulations for unstable particles. Again, Jupiter played an important role in the final state of particles. In this case, 9 particles were registered on the orbit with $a<2\,AU$ and also one particle reached a Phaethon-like orbit.

\section{Escaping material from Ceres - rough estimate}
To roughly estimate if it is possible for material released from Ceres ($a \sim 2.766\,AU$) to get close to the 8:3 MMR with Jupiter ($a \sim 2.705\,AU$), we arranged a simple simulation of particles being released from Ceres in the opposite direction to the heliocentric motion of Ceres, because we needed to obtain smaller values of semi-major axis. We took into account the Sun, all of the planets, and the Moon, and we used the IAS15 integrator to integrate the released particles for $100\,yr$. We found that a particle which was released with 1/15 of the speed of Ceres was able to reach the orbit with $a\sim 2.5\,AU$ and $e\sim 0.19$ (without any other effects). The speed of 1/15 of the instantaneous speed of Ceres represents, approximately, the speed $1.3\,km\,s^{-1}$ at perihelion and $1.1\,km\,s^{-1}$ at aphelion. By adding the escape velocity from the surface of Ceres ($v_{esc} = 0.516\,km\,s^{-1}$) to these speeds, we got to speeds roughly in the range $1.6-1.8\,km\,s^{-1}$. We know that meteorites from Mars and the Moon exist. This means mechanisms which are able to create meteoroids escaping from the surface of Mars ($v_{esc} \sim 5\,km\,s^{-1}$) or the Moon ($v_{esc} \sim 2.4\,km\,s^{-1}$) must exist. Escape velocities from Mars and the Moon are significantly greater than the considered range of speeds. Therefore, we assume that it is possible for particles released from Ceres to get to the 8:3 MMR with Jupiter.

\section{Conclusions}
In this work, we dealt with the transportation of material close to the Earth via the 8:3 MMR with Jupiter. We studied this resonance in combination with dwarf planet Ceres, which could be one of the sources of meteoroids reaching the NEO region. 
We wanted to know if it is possible for material released from Ceres to get close to the Earth via the 8:3 MMR with Jupiter. We applied the FLI map method to the region of 8:3 MMR with Jupiter, to differentiate between stable and unstable orbits. Based on the FLI maps, we integrated 500 of the most stable and 500 of the most unstable particles for $15\,Myr$, for mean anomaly $M=60^\circ$ and also for $M=30^\circ$. In the case of $M=60^\circ$ integrations revealed that stable particles usually evolved in a very peaceful way (Fig.~\ref{stable}).  Only 4 ($0.8\,\%$) particles managed to reach the limit of $q\leq 1.3\,AU$ and therefore became NEOs. On the other hand, in the case of unstable particles, we registered 209 ($41.8\,\%$) particles reaching the orbit with perihelion distance $q \leq 1.3\,AU$ at some point during the simulation.  
For mean anomaly $M=30^\circ$, the evolution of stable particles was significantly different. 238 ($47.6\,\%$) reached an orbit with $q\leq 1.3\,AU$. 
In the case of unstable particles for $M=30^\circ$, we obtained very similar results to unstable particles for $M=60^\circ$. In this case, we again registered 209 ($41.8\,\%$) particles on an orbit with $q\leq 1.3\,AU$, which is the same amount as in the case of $M=60^\circ$. 
The main difference between stable particles for $M=60^\circ$ and $M=30^\circ$ was that in the case of $M=30^\circ$ the majority of test particles had already started with higher eccentricities. 
The results for stable particles indicate that short-term FLI maps are more suitable for finding chaotic orbits, than for detecting the stable ones.
In all cases, except stable particles for $M=60^\circ$, we registered a number of particles that were ejected to hyperbolic orbit or to an orbit with $a>100\,AU$.
These ejections were predominantly caused by Jupiter (for example, Fig.~\ref{init_end_unstable}). By rough estimation, we showed that material released from Ceres is able to get to the 8:3 MMR with Jupiter. Therefore, our results show that 8:3 MMR with Jupiter is capable of transporting material to the NEO region, possibly even material from dwarf planet Ceres.

\section*{Acknowledgements}
We thank the anonymous reviewer for their comments and suggestions, which helped us improve the paper.
This work was supported by the VEGA - the Slovak Grant Agency for
Science, grant No. 1/0596/18, by the Slovak Research and Development
Agency under the contract No. APVV-16-0148 and by the Comenius
University, Slovakia, grant UK/369/2020.

\section*{Data Availability}

No new data were generated or analysed in support of this research.
 



\bibliographystyle{mnras}
\bibliography{references-short2}








\bsp	
\label{lastpage}
\end{document}